\documentclass[a4paper,12pt,english]{article}

\usepackage{amsfonts,bm,amssymb,euscript,array,babel,cite}
\usepackage{mathtools}
\usepackage{tikz-cd}
\usepackage{amsmath}
\usepackage{tensor}
\usepackage{hhline}
\usepackage[amsmath]{empheq}
\usepackage{hyperref}
\usepackage{cancel}
\usepackage{mathrsfs}
\usepackage{pdfsync}
\usepackage{tikz}

\newsavebox{\mysaveboxM}
\newsavebox{\mysaveboxT}

\makeatletter

\newcommand{\dd}{\mathrm{d}}

\newcommand{\w}{\wedge}

\newcommand{\be}{\begin{equation}}
\newcommand{\ee}{\end{equation}}
\newcommand{\sfrac}[2]{{\textstyle\frac{#1}{#2}}}
\def\nn{\nonumber}
\def \bea{\begin{eqnarray}} 
\def\eea{\end{eqnarray}}
\newcommand{\mf}{\mathfrak}
\def\mc{\mathcal}

\def\bi{\begin{itemize}} 
\def\ei{\end{itemize}}

\makeatother

\def\a{\alpha} \def\b{\beta} \def\g{\gamma} \def\G{\Gamma} \def\d{\delta} 
\def\e{\epsilon} 
   
\def\l{\lambda}  \def\m{\mu}
\def\n{\nu} \def\o{\omega}   \def\r{\rho}
\def\s{\sigma} \def\S{\Sigma}

\def\R{{\mathbb R}}   \def\X{\mathbb X}

\def\one{\mbox{1 \kern-.59em {\rm l}}}

\numberwithin{equation}{section}

\begin{document}

\makeatother
\parindent=0cm
\renewcommand{\title}[1]{\vspace{10mm}\noindent{\Large{\bf #1}}\vspace{8mm}} \newcommand{\authors}[1]{\noindent{\large #1}\vspace{5mm}} \newcommand{\address}[1]{{\itshape #1\vspace{2mm}}}

\begin{titlepage}

\begin{flushright}
	RBI-ThPhys-2020-29
\end{flushright}

\begin{center}

\title{ {\large {Gauged sigma-models with nonclosed 3-form \\[5pt] and twisted Jacobi structures}}}

  \authors{\large Athanasios {Chatzistavrakidis}{$^{\a,}$}{\footnote{Athanasios.Chatzistavrakidis@irb.hr}} and Grgur \v{S}imuni\'c{$^{\a,}$}{\footnote{Grgur.Simunic@irb.hr}}
    }
 
 \vskip 3mm
 
  \address{ $^{\a}$ Division of Theoretical Physics, Rudjer Bo\v skovi\'c Institute \\ Bijeni\v cka 54, 10000 Zagreb, Croatia \\

 }

\vskip 2cm

\begin{abstract}

We study aspects of two-dimensional nonlinear sigma models with Wess-Zumino term corresponding to a nonclosed 3-form, which may arise upon dimensional reduction in the target space. Our goal in this paper is twofold. In a first part, we investigate the conditions for consistent gauging of sigma models in the presence of a nonclosed 3-form. In the Abelian case, we find that the target of the gauged theory has the structure of a contact Courant algebroid, twisted by a 3-form and two 2-forms. Gauge invariance constrains the theory to (small) Dirac structures of the contact Courant algebroid. In the non-Abelian case, we draw a similar parallel between the gauged sigma model and certain transitive Courant algebroids and their corresponding Dirac structures. In the second part of the paper, we study two-dimensional sigma models related to Jacobi structures. The latter generalise Poisson and contact geometry in the presence of an additional vector field. We demonstrate that one can construct a sigma model whose gauge symmetry is controlled by a Jacobi structure, and moreover we twist the model by a 3-form. This construction is then the analogue of WZW-Poisson structures for Jacobi manifolds.        

\end{abstract}

\end{center}

\vskip 2cm

\end{titlepage}

\setcounter{footnote}{0}
\tableofcontents


\section{Introduction}
\label{sec1}

The Poisson sigma model  is a two-dimensional topological quantum field theory whose target space is a Poisson manifold \cite{SchallerStrobl,Ikeda}. It features a number of attractive properties and applications, the most well-known being that its path integral quantization (when viewed as an open string theory on a disk) provides a physical interpretation of the Kontsevich formula for deformation quantization of Poisson manifolds and the noncommutativity of string endpoint coordinates  \cite{CattaneoFelder1,CattaneoFelder2}. Furthermore, the model may be directly related to two-dimensional gravity with or without torsion \cite{Klosch:1995fi,Grumiller:2002nm}, to the completely gauged topological Wess-Zumino-Witten (WZW) models \cite{Alekseev:1995py} and to the A and B models \cite{Alexandrov:1995kv}. We refer to \cite{Ikeda:2012pv} for further developments and literature. 

There exist two interesting generalizations of the Poisson sigma model, which will be relevant in the present paper. The first generalization is to Dirac sigma models, first introduced in Ref. \cite{Kotov:2004wz} as interpolations between the Poisson sigma model and $G/G$ WZW models. They owe their name to the fact that they correspond to target spaces being Dirac manifolds \cite{dirac}, a concept that includes Poisson and presymplectic as special cases. The study of Dirac manifolds is where the Courant bracket naturally appears, whose properties are systematically encoded in the notion of a Courant algebroid \cite{Liu:1995lsa}, all these being central concepts in generalized geometry \cite{gg1} and its applications in physics, notably in string theory. 

On the other hand, Dirac sigma models bear a strong relation to gauging models of strings in backgrounds of a metric $G$ and a closed 3-form $H$. Starting with the early work of \cite{hs1,hs2}, the gauging of sigma models with Wess-Zumino term has been revisited in recent years in light of the developments in generalized geometry, and it has been realised that Dirac structures are strongly related to consistency of the gauging procedure \cite{Salnikov:2013pwa,Plauschinn:2013wta,ChatzistavrakidisAHP,Chatzistavrakidis:2016jci,Severa:2019ddq}. 
See also \cite{Ikeda:2018rwe,Ikeda:2019pef} for an interesting Hamiltonian perspective. It is this feature that we wish to study further in this paper.
To be specific, the 3-form $H$ in all above studies is assumed closed, a property that plays a crucial role in the analysis and in particular in the derivation of the constraints that gauge invariance imposes to the theory. However, it can happen that the 3-form is not closed. Ultimately, this is the case in heterotic string theory, where the 3-form acquires $\alpha'$ corrections. Here we will be interested in simpler situations and comment on the heterotic case in the concluding section. Nonclosed 3-forms may also arise upon dimensional reduction in the target space, for instance when it has the structure of a circle fibration \cite{BouwknegtLectures,Coimbra:2014qaa} and more generally in reduction of Courant algebroids \cite{Severa:2015hta,Severa:2017oew,Bursztyn:2005vwa,RCA,Wright}. 

Motivated by the above, in Section \ref{sec2} we study the gauging of two-dimensional sigma models in the presence of a nonclosed 3-form that can arise upon dimensional reduction as mentioned above. The gauging is performed along vector fields that leave invariant the background fields of the theory and satisfy, in general, a non-Abelian Lie algebra.{\footnote{We do not consider here the more general setting of \cite{ChatzistavrakidisAHP,ks1}, where strict invariance can be relaxed.}} We examine separately the cases when there is a single extra dimension in the theory before dimensional reduction (we call this case Abelian) and when there are multiple extra dimensions (non-Abelian case.) In the Abelian case, we find that gauge invariance of the gauged theory imposes additional constraints, as expected from the corresponding results with closed 3-form. These constraints acquire an elegant geometric interpretation when one considers the structure of a contact Courant algebroid over the extended vector bundle $TM\oplus\R\oplus\R\oplus T^{\ast}M$ twisted by three tensors, specifically a 3-form $H$ and two 2-forms $\Omega$ and $F$ whose product controls the 4-form $\dd H$ \cite{BouwknegtLectures,Wright}; they are identified with conditions for Dirac structures of the twisted contact Courant algebroid. The same geometric interpretation persists in the more general, non-Abelian case, with the difference that the corresponding higher structure is a nonexact Courant algebroid of the type described in \cite{RCA}, over an extended bundle $TM\oplus\mc{G}\oplus T^{\ast}M$, with $\mc{G}$ a bundle of quadratic Lie algebras.    
 
 The second interesting generalization of the Poisson sigma model that we would like to discuss,{\footnote{To avoid confusion, we mention that as long as Dirac sigma models contain a Wess-Zumino term, this generalization is included in them. However, conceptually it is different and it is one that precedes the construction of Dirac sigma models. Here we follow an order of presentation more suitable for the flow of the present paper.}} is the so-called Wess-Zumino or $H$-twisted Poisson sigma model introduced in \cite{Klimcik:2001vg}, and the associated WZW-Poisson manifolds \cite{SeveraWeinstein}. These are obtained by adding a topological 3-form term to the Poisson sigma model and realising that gauge invariance under an extension of its characteristic symmetries can be maintained as long as the 3-form controls in a specific way the nonvanishing of the Schouten bracket of the quasi-Poisson structure with itself. Note that the BV and BFV formulations of this model were found only recently \cite{Ikeda:2019czt}.

Motivated by the Poisson sigma model and the existence of its $H$-twisted extension, our second goal is to investigate from the viewpoint of gauging whether similar structures exist for Jacobi manifolds. Jacobi structures and the associated manifolds are a natural generalization of their Poisson counterparts \cite{Jacobi}. A simple way to introduce them is the observation that if a Poisson bivector $\mc P$ is multiplied by a smooth function $f$, then the resulting structure is not Poisson any longer, unless the Hamiltonian vector field of $f$ vanishes \cite{Crainic}. The resulting structure is called Jacobi and it is given by a bivector $\Pi$ along with a vector field $V$ satisfying certain conditions to be presented below. Jacobi structures are rigid under multiplication by smooth functions. Alternatively, recalling that a Poisson bracket endows the space of functions with a Lie algebra structure and it satisfies the Leibniz rule, the Jacobi bracket relaxes the latter property to just being an operation of the local type \cite{Vaisman}. 

Vector fields on manifolds, Poisson and odd-dimensional contact manifolds are all examples of Jacobi manifolds. Interestingly, there exists a Poissonization of a Jacobi manifold $(M,\Pi,V)$, meaning that one may construct a Poisson structure $\widetilde{\cal P}$ on $M\times \R_{+}$ out of $\Pi$ and $V$, see e.g. \cite{Crainic}. This is the analog of dimensional oxidation in physics parlance (or, inversely seen, dimensional reduction,) and it will be useful in Section \ref{sec3}. There we construct a two-dimensional sigma model with target space a Jacobi manifold, as a natural generalization of the Poisson sigma model. Indeed, we show that the action functional of the model is gauge invariant under certain infinitesimal gauge transformations provided that the characteristic conditions that define a Jacobi structure are satisfied. Moreover, we study the gauge algebra and show that it is an open one, namely it closes only on-shell, as expected. Using Poissonization, we discuss the relation of the Jacobi sigma model to the Poisson sigma model in one target space dimension higher. Furthermore, we consider the extension of the action functional of a Jacobi sigma model by Wess-Zumino terms. This is similar in spirit to the $H$-twisted Poisson sigma model. We show that there are two such terms, corresponding to 3-form and 2-form twists in the defining relations of a Jacobi structure. The gauge structure of the corresponding field theory is shown to be consistent with the twisted Jacobi structure of \cite{twistedJacobi}.    

\section{Gauging nonlinear sigma models with nonclosed 3-form}
\label{sec2} 

\subsection{Preliminaries}
\label{sec21}

The propagation of strings in target spacetimes $M$ is described by nonlinear sigma models. 
These are two-dimensional field theories on a Riemann surface $\S_2$, the worldsheet.{\footnote{For our purposes, we restrict in this section to  worldsheets at the lowest order in the string perturbation expansion.}} The basic fields of the theory are $\text{dim}\,M\equiv d$ scalars $X^i$, the components of the sigma model map $X: \S_2\to M$. The nonlinear couplings of the theory correspond to target space background fields, which are typically pull-backs of geometric data defined on $M$. In the simplest closed bosonic string case, these are a (pseudo-)Riemannian metric $G$ and a 2-form, the Kalb-Ramond field $B$, or more generally its curvature $H=\dd B$, which is a closed 3-form.{\footnote{In addition, there is also the scalar dilaton field, which we will not consider further in this paper.}} When written in terms of $H$, the sigma model contains a Wess-Zumino term which is supported on a three-dimensional source manifold $\Sigma_3$ with boundary the worldsheet $\Sigma_2$. In that case, the map $X$ is extended to one from $\S_3$ to $M$. Moreover, the 3-form $H$ does not have to be exact; however, as long as it defines an integral cohomology class, its Wess-Zumino term does not depend on the choice of $\Sigma_3$ and although it is ambiguous up to an integer constant, the corresponding path integral is not ambiguous \cite{Witten:1983ar}. It is in this sense that we will deal with Wess-Zumino terms in the following.

In this paper, we are interested in a more general situation, where the 3-form $H \in \Omega^{3}(M)$ is not closed. Since its exterior derivative $\dd H$ is a 4-form, we assume that $M$ is also equipped with two 2-forms, say $\Omega$ and $F$, such that 
\be 
\dd H=- \langle F\w \Omega \rangle\,,\label{Hcondition}
\ee 
where the angle brackets denote the fact that the 2-forms may be also valued in additional bundles; for example they can have a gauge index, as will be the case here, in which case a trace should be taken in the right-hand side. In this general setting, we therefore consider $F,\Omega \in \Omega^{2}(M;\mc G)$, where $\mc G$ is a bundle of quadratic Lie algebras where the two 2-forms take values. This means that $\mc G$ is equipped with a nondegenerate, symmetric inner product $\langle\,\cdot,\cdot\,\rangle_{\mc G}$, which is ad-invariant, essentially a $\mc G$-metric.{\footnote{More precisely a pseudo-metric, but we will simply refer to it as metric here.}} 

As mentioned in the introduction, such cases can arise upon dimensional reduction in the target space. Specifically, let us assume that the target has the structure of a circle fibration with $d$-dimensional base $M$ and $S^1$ fibers and it is equipped with a metric $\widetilde{G}$ and a \emph{closed} 3-form $\widetilde{H}$. The sigma model with Wess-Zumino term in this target space is 
\begin{equation}\label{Stilde}
\widetilde{S}_0=-\int_{\Sigma_2}\frac{1}{2}\widetilde{G}_{\mu\nu}(\widetilde{X})\dd\widetilde{X}^\mu\wedge\ast\dd \widetilde{X}^\nu-\int_{\Sigma_3}\frac{1}{3!}\widetilde{H}_{\mu\nu\sigma}(\widetilde{X})\dd\widetilde{X}^\mu\wedge\dd\widetilde{X}^\nu\wedge\dd\widetilde{X}^\sigma\,,
\end{equation}	
where $\widetilde{X}$ is the sigma model map to the $(d+1)$-dimensional target space. Background fields are pull-backs in the sense that e.g. $\widetilde{G}_{\m\n}(\widetilde X)=\widetilde X^{\ast}G_{\m\n}(\widetilde x)$, where $\widetilde x^{\m}$ are coordinates on the target space.  We use differential form notation, where the metric of the worldsheet is hidden in the Hodge star operator, and we assume that $\S_2$ has Lorentzian signature. To proceed with dimensional reduction, we make the further assumption that the background fields are invariant under the isometry generated by the Killing vector field on $S^{1}$. In adapted coordinates $\widetilde{x}^{\m}=(x^{i},\Phi)$ where the Killing vector is simply $\partial/\partial\Phi$, this assumption simply translates to the metric and 3-form being independent of $\Phi$. This corresponds to a Kaluza-Klein reduction,{\footnote{Note that in more general cases one could in principle consider Scherk-Schwarz reductions, where the higher-dimensional fields depend on the internal coordinates but the dependence drops out in the lower-dimensional theory due to the symmetry structure of the higher-dimensional one. Such cases are beyond the scope of the present paper.}} where the Ansatz for the metric $\widetilde{G}$ takes the characteristic form (in terms of the line element)
\be \label{KKmetric}
\dd \widetilde{s}^2=G_{ij}(x)\dd x^i\dd x^j+G_{\Phi\Phi}(x)(\dd \Phi+\mf{a})^2\,,
\ee  
with $\mf a=\mf a_i\dd x^i$ being the Kaluza-Klein vector with field strength $F=\dd\mf a$.  In the usual Kaluza-Klein parametrization, $G_{\Phi\Phi}$ is the exponential of a scalar field, however we keep the discussion general here. Furthermore, the 3-form is decomposed accordingly as 
\be \label{KKH}
\widetilde{H}=H+\Omega\w (\dd \Phi+\mf{a})\,,
\ee 
where the 3-form $H$ and the 2-form $\Omega$ depend only on the lower-dimensional spacetime, i.e. they are independent of $\Phi$. Then, given that $\dd \widetilde{H}=0$, one finds that the lower-dimensional 3-form is not closed but instead satisfies 
\be 
\dd H=-\Omega\w F\,,
\ee 
whereas $\Omega$ is itself a closed 2-form, $\dd \Omega=0$. This is then how \eqref{Hcondition} is obtained upon dimensional reduction. With this starting point, \cite{Coimbra:2014qaa} studied the generalised geometry of the lower-dimensional target space theory. Instead, our purpose is to study the corresponding sigma model with Wess-Zumino terms from the lower-dimensional viewpoint and its gauging.

In the above set up, one may then consider as starting point the following action functional:
  \bea 
  S_{0}= S_{0,\text{kin}} + S_{0,\text{WZ}}\,,
  \label{abelianungauged}
  \eea 
 split into a kinetic sector and one that contains Wess-Zumino terms, corresponding precisely to the Kaluza-Klein reduction Ansatz \eqref{KKmetric} and \eqref{KKH} respectively: 
 \bea 
 \label{S0kin} S_{0,\text{kin}}&=&-\int_{\S_2}\, \left(\frac 12 G_{ij}(X)\dd X^{i}\w\ast\dd X^{j}
 +\frac 12 G_{\Phi\Phi}(X)\Lambda\wedge \ast\Lambda\right)\,,\\
 S_{0,\text{WZ}}&=&-\int_{\S_3}\,\left(\frac 1{3!}H_{ijk}(X)\dd X^{i}\w\dd X^{j}\w\dd X^{k}+\frac 1{2!}\Omega_{ij}(X)\dd X^{i}\w\dd X^{j}\w \Lambda\right)\,.
 \eea 
 Here, $\Lambda$ is the (pull-back of the) 1-form
 $ 
  \Lambda=\dd\Phi +\mf{a},
  $   
 and $\Phi(\sigma)$ is the additional scalar corresponding to the reduction picture explained above, which is a function of the local coordinates on the worldsheet. Note that none of the components of the background fields depend on this additional field.

The action \eqref{Stilde} can have target space symmetries generated by a set of vector fields $\widetilde\rho_a=\rho_a^{\mu}(\widetilde X)\partial_{\m}$ that satisfy a non-Abelian Lie algebra 
  \be \label{rhobracket}
  [\widetilde\r_a,\widetilde\r_b]=C_{ab}{}^{c}\widetilde\rho_c\,.
  \ee
  Typically, $C_{ab}{}^{c}$ are structure constants; however, in general they may be $\widetilde X$-dependent structure functions, as explained for instance in \cite{ChatzistavrakidisAHP}, in which case one can have a Lie algebroid instead.{\footnote{Recall that a Lie algebroid is a vector bundle $E$ over $M$ with a (anchor) map $\rho:E\to TM$ and a Lie algebra structure $[\cdot,\cdot]$ on the sections $\Gamma(E)$ that satisfies the Leibniz rule $[e_1,fe_2]=f[e_1,e_2]+\rho(e_1)f\cdot e_2$.}} In the present section, this statement will be somewhat neglected; from now on, one may safely assume that $C_{ab}{}^{c}$ are constants, albeit keeping in mind that this can be generalized in a straightforward way. In the reduction spirit employed here, these vector fields may be decomposed as 
  \be 
\widetilde{\r}_a=\rho_a^{i}(X)\partial_i+ \widetilde{f}_a(X)\partial_{\Phi}\,,
\ee
where $\widetilde{f}_a$ are functions on $M$ (see also Section 5.3.6 of \cite{BouwknegtLectures}.) This leads to a set of vector fields $\rho_a$ satisfying the algebra 
\be 
[\r_a,\r_b]=C_{ab}{}^{c}\rho_c\,,
\ee 
with the same structure constants as before. Moreover, it immediately follows that
 \be \label{tildef}
2\iota_{\rho_{[a}}\dd\widetilde{f}_{b]}=C_{ab}{}^c\widetilde{f}_c\,.
\ee 
In the following, we consider another parametrization for these functions, specifically 
\be 
\widetilde{f}_{a}=f_a-\iota_{\rho_{a}}\mf a\,,
\ee 
in which case \eqref{tildef} becomes an equation for the functions $f_a$ and reads  
\be \label{fbracket}
2\iota_{\rho_{[a}}\dd f_{b]}=C_{ab}{}^{c}f_c-\iota_{\rho_a}\iota_{\rho_b}F\,,
\ee 
 where  $F=\dd \mf a$ is the Abelian field strength of the 1-form $\mf a$.
 The symmetries generated by $\r_a$ for the action functional \eqref{abelianungauged} manifest themselves upon considering the following transformations for the scalar fields $X^{i}$ and $\Phi$: 
  \bea 
\label{deltaX}  \d X^{i}&=&\rho_a^{i}\epsilon^{a}\,,\\
\label{deltaPhi}  \d \Phi &=&\widetilde{f}_a\e^a\,,
  \eea 
  where $\e^{a}$ are rigid symmetry parameters.
   Moreover, for future reference, it is useful to calculate the transformation of the 1-form $\Lambda$, which turns out to be 
  \be 
  \d \Lambda=\dd(f_a\e^{a})+\iota_{\rho_a} F\,\e^{a}\,,
  \ee  
  We observe that in reference to $\Lambda$, the combination $f_a\e^{a}$ behaves as a single transformation parameter, and moreover the second 2-form $F$ appears in the transformation rule. Note that in the Abelian case, both $F$ and $\Omega$ are closed, namely $\dd F=0=\dd\Omega$.
  
  One can now compute the transformation of the action $S_0$ under \eqref{deltaX} and \eqref{deltaPhi} for constant transformation parameters $\e^{a}$ and examine under which conditions the action is invariant. Taking into account that $\dd H=-F\w\Omega$, one can show that this is true if and only if 
  \bea
&&{\cal L}_{\rho_a}G=0\,, \qquad {\cal L}_{\rho_a}G_{\Phi\Phi}=0\,, \qquad \iota_{\rho_a}F=-\dd f_a\,,\label{metricconditions} \\[4pt]
 && \iota_{\rho_a}H+f_a\Omega-g_aF=\dd\theta_a\,, \qquad   \iota_{\rho_a}\Omega=\dd g_a\,,  \label{rigidconditionsabelian}
  \eea  
   where $\theta_a$ and $g_a$ are arbitrary 1-form and function respectively. Note that $\rho_a$ are Killing vector fields for the metric $G=G_{ij}(x)\dd x^i\otimes \dd x^j$ of the Kaluza-Klein Ansatz \eqref{KKmetric}. However, one should keep in mind that this is not the full nonlinear coupling in  the kinetic term of the scalar fields $X$. As can be seen from \eqref{S0kin}, the components of the full metric should be identified with $G_{ij}+G_{\Phi\Phi}\mf a_i\mf a_j$, i.e. the usual Kaluza-Klein metric in the lower-dimensional space. The vectors $\rho_a$ are not isometries of this metric though. This is expected in view of the higher-dimensional origin of these conditions; indeed, the higher-dimensional vector 
  $\widetilde\rho_a$  is a Killing vector for $\widetilde{G}$.

  A similar discussion follows for the non-Abelian case, where $\Omega^{\a}$ and $F^{\a}$ are $\mc G$-valued and there is a worth of $\text{dim}\,\mc G$ 1-forms $\Lambda^{\a}=\dd\Phi^{\a}+\mf a^{\a}$, where $\a$ is a gauge index. The transformation rule of the scalars $X^{i}$ remains the same as in \eqref{deltaX}, whereas the one of $\Lambda^{\a}$ reads as 
  \be \label{deltaVa}
  \d \Lambda^{\a}=\dd (f_a^{\a}\e^{a})+\iota_{\rho_a}\dd\mf a^{\a}\e^{a}\,,
  \ee  
   Note that in our conventions, the non-Abelian field strength is given as 
  \be 
  F^{\a}=\dd \mf a^{\a}-\frac 12 K^{\a}{}_{\b\g}\mf a^{\b}\w\mf a^{\g}\,,
  \ee 
  where $K^{\a}{}_{\b\g}$ are the structure constants of $\mc G$; however, it should be noted that only $\dd\mf a^{\a}$ appears in the transformation of $\Lambda^{\a}$. The ungauged action functional has the form 
   \bea 
  S_{0}&=&S_{0,\text{kin}} -\int_{\S_3}\,\left(\frac 1{3!}H_{ijk}(X)\dd X^{i}\w\dd X^{j}\w\dd X^{k}+\frac 1{2!}\Omega_{\a ij}(X)\dd X^{i}\w\dd X^{j}\w \Lambda^{\a}\right) \nn \\ && \quad -\int_{\S_3} \left(-\frac 12 K_{\a\b\g}\Lambda^{\a}\w \Lambda^{\b}\w \mf a^{\g}+\frac 1{3!}K_{\a\b\g}\Lambda^{\a}\w \Lambda^{\b}\w \Lambda^{\g}\right)\,,\label{nonabelianungauged}
  \eea 
  where we use the same symbol $S_0$ as in the Abelian case, since the latter is simply obtained from the non-Abelian one in an obvious way. Note that the gauge indices are raised and lowered with the $\mc G$-metric (essentially the Killing form), which in a basis of Lie algebra generators $\{T^{\a}\}$ we denote as $k^{\a\b}$. 
  Then the conditions \eqref{rigidconditionsabelian} generalize in the non-Abelian case to
  \begin{subequations}\label{conditions full}\bea && \iota_{\rho_a}H+f_a^{\a}\Omega_{\a}-g_{a\a}\dd\mf a^{\a}-\frac 12 K_{\a\b\g}\widetilde{f}_{a}^{\g}\mf a^{\a}\w\mf a^{\b}=\dd\theta_a\,, \\ && \iota_{\rho_a}\left(\Omega_{\a}-\sfrac 12 K_{\a\b\g}{\mf a}^{\b}\w\mf a^{\g}\right)=\dd g_{a\a}\,,
  	\\&&  \iota_{\rho_a}\left(F^{\a}+\frac 12 K^{\a}{}_{\b\g}\mf a^{\b}\w\mf a^{\g}\right)=-\dd f_{a}^{\a}\,,
  \eea\end{subequations} 
  provided that 
  \be 
  \dd \Omega_{\a}+K_{\a\b\g}\mf a^{\b}\w\dd\mf a^{\g}=0
  \ee 
  and that $K_{\a\b\g}\widetilde{f}^{\g}_a$ is constant.
  These conditions and their interpretation will be revisited in more detail in the gauged version of the theory.

  \subsection{Gauging with nonclosed 3-form}
  \label{sec22} 
  
  The global symmetries generated by the vector fields 
  $\rho_a+\widetilde f_a\partial_{\phi}$ 
  can be promoted to local symmetries of the action upon considering transformation parameters $\e^{a}=\e^{a}(\s)$ that depend on the worldsheet coordinates. This allows us to consider gaugings of the action \eqref{abelianungauged} along the foliation generated by the corresponding vector fields. By gauging we mean finding a new action functional $S$ that depends on additional 1-form gauge fields $A$ coupled to the theory, such that when $A$ are set to zero one obtains the original ungauged action $S_0$ and in addition $S$ possesses a gauge symmetry associated to the vector fields.{\footnote{See \cite{ChatzistavrakidisAHP} for more refined definitions, including the concept of strict gauging which we will not explore further here.}}  
  
  According to the above, we consider additional worldsheet 1-forms $A^{a}$ that we wish to couple to the theory. These gauge fields take values in some gauge bundle $\mc E$, in which we may consider a local basis of sections $e_a$ such that $A=A^{a}e_a$.  $\mc E$ is taken to be a Lie algebroid,
  with a bundle map $\rho$ to the tangent bundle of $M$ such that $\r(e_a)=\r_a$. The requirement of a Lie algebroid means that the map $\rho$ is a homomorphism of bundles. As such, if the bracket operation on $\mc E$ satisfies 
  \be 
  [e_a,e_b]_{\mc E}=C_{ab}{}^{c}e_c\,,
  \ee 
  then \eqref{rhobracket} follows. For the purposes of the present paper, we consider that $\mc E$ is a direct sum bundle $\mc E=L\oplus L'$ over $M$, for some bundles $L$ and $L'$. For instance, this allows us to consider Lie algebroids such as $TM\oplus \R$ and $T^{\ast}M\oplus \R$ in the Abelian case; the fact that these are indeed Lie algebroids is discussed for example in \cite{Iglesias} (see also \cite{GM}). 
  
  As typical for sigma model actions with Wess-Zumino term, gauging proceeds by a nonminimal coupling of the gauge fields $A^{a}$ to the topological sector of the theory, while they are coupled minimally to the kinetic sector. This is facilitated by the following candidate gauged action functional 
  \bea 
  S&=&S_{\text{kin}}-\int_{\S_2}\, \left(A^a\w\theta_a+g_{a\a}A^{a}\w \Lambda^{\a}+\frac 12 \g_{ab}A^a\w A^b\right) \nn\\ && \quad -\int_{\S_3}\,\left(\frac 1{3!}H_{ijk}(X)\dd X^{i}\w\dd X^{j}\w\dd X^{k}+\frac 1{2!}\Omega_{\a ij}(X)\dd X^{i}\w\dd X^{j}\w \Lambda^{\a}\right) \nn \\ && \quad -\int_{\S_3} \left(-\frac 12 K_{\a\b\g}\Lambda^{\a}\w \Lambda^{\b}\w \mf a^{\g}+\frac 1{3!}K_{\a\b\g}\Lambda^{\a}\w \Lambda^{\b}\w \Lambda^{\g}\right)\,,\label{nonabeliangauged}
  \eea 
  where $\theta_a=\theta_{ai}\dd X^{i}$, $g_{a\a}$, $\g_{ab}$ are arbitrary 1-form and functions of $X$ respectively. The gauge fields $A^{a}$ transform as is typical for a nonlinear gauge theory\footnote{This is certainly not the most general transformation for the gauge field in this context, but it is the one considered in the original papers \cite{hs1,hs2}. It corresponds to the standard case where the gauge symmetry originates from a corresponding global symmetry, which we also assume here. More generally, one can examine whether a gauge theory exists even without an underlying global symmetry, by adding terms proportional to $DX^{i}$ and $\ast DX^{i}$ to the transformation rule, essentially introducing two vector bundle connections on the Lie algebroid, see e.g. \cite{Chatzistavrakidis:2016jci,ChatzistavrakidisAHP}. We do not examine this apparently more general situation here. However, this does not affect the main result which is presented below, since in both cases one obtains conditions for Dirac structures of a Courant algebroid. The only difference regards the background fields, which in the present case are assumed invariant, while in the general case this assumption is relaxed.} 
  \be \label{deltaA}
  \d A^{a}=\dd\e^{a}+C^{a}{}_{bc}A^{b}\e^{c}\,.
  \ee   
  Moreover, the gauged kinetic sector takes the form 
  \be 
  S_{\text{kin}}=-\int_{\S_2}\frac 12 G_{ij}DX^{i}\w\ast DX^{j}+\frac 12 G_{\a\b}\hat{\Lambda}^{\a}\w\ast\hat{\Lambda}^{\b}\,,
  \ee 
  where the worldsheet differential $DX$ and the 1-form $\hat{\Lambda}$ are defined as 
  \bea DX^{i}&=&\dd X^{i}-\r^{i}_{a}A^{a}\,,\\
  \hat{\Lambda}^{\a}&=&D\Phi^{\a}+\mf a_i^{\a}DX^{i}\,, \eea
  with $D\Phi^{\a}=\dd\Phi^{\a}-\widetilde{f}_a^{\a} A^{a}$. They are both covariant as worldsheet 1-forms, since they transform as 
  \bea 
  \d DX^{i}=\partial_j\rho^{i}_{a}\epsilon^{a}DX^{i} \quad \text{and} \quad \d\hat{\Lambda}=(\partial_if_b+(\iota_{\rho_{b}}\dd\mf a)_i)\epsilon^{b}DX^{i}\,. 
  \eea 
   Note that in principle one can also consider additional equation of motion symmetries in the transformation of $A^{a}$, also called trivial gauge transformations \cite{HT}, which are important when the gauge algebra of the model is open. In the present section we consider only gauge algebras that close off shell, but in Section \ref{sec3} we will encounter also open algebras and we will then take into account such additional gauge symmetries.  
  
  We now examine under which conditions the action $S$ is invariant under the above gauge transformations. As for the kinetic sector, this is separately gauge invariant provided that \eqref{metricconditions}, or the corresponding non-Abelian extension of them, hold. On the other hand, the topological sector imposes additional constraints. In order to cancel all terms supported on $\S_3$, we take into account the following identities 
  \begin{subequations}\label{dtwists}\begin{align}
  \dd \mc H&=-F^{\a}\w\Omega_{\a}\,,\\
  \dd F^{\a}&=-K^{\a}{}_{\b\g}\dd \mf a^{\b}\w\mf a^{\g}\,,\\
  \dd \Omega^{\a}&=K^{\a}{}_{\b\g}\dd\mf a^{\b}\w\mf a^{\g}\,,
  \end{align}\end{subequations}
  where we have defined the improved 3-form 
  \be 
  {\mc H}=H-\frac 1{3!}K_{\a\b\g}\mf a^{\a}\w\mf a^{\b}\w\mf a^{\g}\,.
  \ee 
  Note that although $F$ and $\Omega$ seem related in view of \eqref{dtwists}, we have treated them separately because in the Abelian case they can be independent. 
   Then it is straightforward to show that the topological sector of the action $S$ transforms as 
  \bea 
\d S_{\text{top}}&=&-\int_{\S_2}\left(\g_{ab}-\r_a^{i}\theta_{bi}-f_{a}^{\a}g_{b\a}\right)\dd\e^{a}\w A^{b}\nn \\[4pt]
&  -&\int_{\S_2}\e^{a}\left[\left(\frac 12 \Omega_{\a ij}f_{a}^{\a}+\frac 12 \r_a^{k}H_{ijk}-g_{a\a}\partial_{i}\mf a_{j}^{\a}-\partial_{i}\theta_{aj}\right)\dd X^{i}\w\dd X^{j}\right. \nn\\[4pt] 
&& \quad \quad\left.+ \left(-\partial_ig_{a\a}
+\r_a^{j}\Omega_{\a ji}-K_{\a\b\g}f_a^{\b}\mf a_{i}^{\g}\right)\dd X^{i}\w \Lambda^{\a}+ \frac 12 K_{\a\b\g}\left(f_{a}^{\g}-\r_a^{i}\mf a_{i}^{\g}\right)\Lambda^{\a}\w \Lambda^{\b}\right. \nn\\[4pt]
&& \quad \quad \left. + \left(C_{ab}^{c}\theta_{ci}-\r_{a}^{j}\partial_{j}\theta_{bi}-\theta_{bj}\partial_{i}\r_{a}^{j}-g_{b\a}\partial_if_{a}^{\a}-g_{b\a}(\iota_{\r_a}\dd\mf a^{\a})_i\right)\dd X^{i}\w A^{b}\right.\nn\\[4pt]
&&\quad \quad +\left.\left(C_{ab}^{c}g_{c\a}-\r_a^{i}\partial_ig_{b\a}\right) \Lambda^{\a}\w A^{b}+\left(\frac 12 \r^{i}_{a}\partial_i\g_{bc}-C_{ab}^{d}\g_{dc}\right)A^b\w A^c\right]\,,\label{trafoS}
  \eea 
  where we have also used the Jacobi identity for the structure constants $K_{\a\b\g}$.
Requiring invariance of the action thus leads to the following conditions on the background fields, 
\begin{subequations}\label{conditions}\begin{align}
\label{condition1}\iota_{\r_a}H+f_{a}^{\a}\Omega_{\a}-g_{a\a}\dd\mf a^{\a}&=\dd\theta_a\,,
\\[4pt]
\label{condition2}\iota_{\r_a}\Omega_{\a}-K_{\a\b\g}f_{a}^{\b}\mf a^{\g}&=\dd g_{a\a}\,\\[4pt]
K_{\a\b\g}\widetilde{f}_a^{\g}&=0\,.\label{condition3}
\end{align}\end{subequations} 
Note that in comparison to the conditions found in the rigid case, $K_{\a\b\g}\widetilde{f}^{\g}_{a}$ is zero rather than constant because $\epsilon$ is not a rigid parameter. This affects only the non-Abelian case, since in the Abelian case $K_{\a\b\g}=0$ anyway. In the non-Abelian case it implies a relation between the functions $f_{a}^{\a}$ and the contraction of the fundamental gauge field $\mf{a}^{\a}$ with the vector fields $\rho_a$. We comment further on this below. 
In addition, there are three constraints that must be satisfied in order to obtain a consistent gauge theory. First, we note from the first term in the variation of the action that 
\be 
\gamma_{ab}=\iota_{\r_a}\theta_{b}+k_{\a\b}f_a^{\a}g_{b}^{\b}\,.
\ee 
Then the three constraints read as 
\begin{subequations}\label{constraints}\begin{align}
\label{constraint1} \g_{(ab)}&=0\, \\
\label{constraint2} {\cal L}_{\rho_a}\theta_b&=C_{ab}^{c}\theta_c-g^{\a}_b\left(\dd f_{a\a}+\iota_{\rho_a}F_{\a}+K_{\a\b\g}f_a^{\b}\mf a^{\g}\right)\,, \\ 
\label{constraint3} \iota_{\rho_a}\dd g_b^\alpha&=C_{ab}^{c}g_c^{\a}\,.
\end{align}\end{subequations}
  Finally, there is a further requirement due to the very last, quadratic in $A$s, term in \eqref{trafoS}, 
  \be 
  {\cal L}_{\rho_a}\gamma_{bc}=C_{a[c}^{d}\gamma_{b]d}\,,
  \ee 
  which, however, is identically satisfied once the previous conditions are taken into account.  
  
  Summarizing the findings of this section, the gauged action functional \eqref{nonabeliangauged} is invariant under the infinitesimal transformations \eqref{deltaX}, \eqref{deltaVa} and \eqref{deltaA} if and only if the conditions \eqref{conditions} hold and the constraints \eqref{constraints} are satisfied, and at the same time the vector fields $\rho_a$ generate isometries for the metric $G$, respectively the higher-dimensional vector fields $\rho_a+\widetilde{f}_{a}^{\a}\partial_{\phi^{\a}}$ generate isometries for the metric $\widetilde{G}$. Our next goal is to understand the geometric meaning of these constraints. 
  
  \subsection{The Abelian case and contact Courant algebroids}
  \label{sec23}
  
  Let us first investigate in more detail the Abelian case where the bundle of quadratic Lie algebras is  ${\mc G}=M\times \R\oplus \R$, in which case $K_{\a\b\g}=0$. The gauged action functional \eqref{nonabeliangauged} then contains only the terms appearing in the first two lines.     
   Moreover, from \eqref{dtwists} we learn that both $F$ and $\Omega$ are closed 2-forms, while the exterior derivative of the 3-form $H$ is the opposite of their wedge product. Note that nothing necessitates any further relation between the two 2-forms in the present case. Moreover, the third of conditions \eqref{conditions} is identically satisfied and does not play any role, in particular it does not relate $f_{a}$ with $\iota_{\rho_{a}}\mf a$. 
   
   We proceed in analyzing the three constraints \eqref{constraints}, taking into account the first two conditions \eqref{conditions}.  First, the constraint \eqref{constraint3} may be rewritten as 
   \bea
    C_{ab}^{c}g_{c}=\rho_{a}^{i}\partial_ig_b=2\rho^{i}_{[a}\partial_{|i|}g_{b]}+\rho_b^{i}\partial_ig_a\overset{\eqref{condition2}}=2\rho^{i}_{[a}\partial_{|i|}g_{b]}+\rho^{i}_b\rho^{j}_a\Omega_{ji}\,,
   \eea 
   or 
   \be \label{constraint3b}
   C_{ab}^{c}g_c=2\iota_{\rho_{[a}}\dd g_{b]}-\iota_{\rho_a}\iota_{\rho_b}\Omega\,,
   \ee 
   where (anti)symmetrization is taken with weight 1 and vertical bars denote exclusion from it.
  This equation looks like closure of a bracket operation and indeed we are going to interpret it as such. Turning to the constraint \eqref{constraint2}, a similar rewriting is possible as follows: 
  \bea 
  C_{ab}^{c}\theta_c&=&{\cal L}_{\rho_a}\theta_b+g_b\dd f_a+g_b\iota_{\rho_a}F \nn\\ 
  &=& 2{\cal L}_{\rho_{[a}}\theta_{b]}-\dd\iota_{\rho_{[a}}\theta_{b]}+\sfrac 12 \,\dd\left(\iota_{\rho_a}\theta_b+\iota_{\rho_b}\theta_a\right)+\iota_{\rho_b}\dd\theta_a+g_b\dd f_a+g_b\iota_{\rho_a}F \nn\\ 
  &\overset{\eqref{condition1}}=& 2{\cal L}_{\rho_{[a}}\theta_{b]}-\dd\iota_{\rho_{[a}}\theta_{b]}+ \dd\iota_{\rho_{(a}}\theta_{b)}-\iota_{\rho_a}\iota_{\rho_b}H+2g_{[b}\iota_{\rho_{a]}}F+g_b\dd f_a+f_a\iota_{\rho_b}\Omega\nn\\ 
  &\overset{\eqref{condition2}}=&2{\cal L}_{\rho_{[a}}\theta_{b]}-\dd\iota_{\rho_{[a}}\theta_{b]}-\iota_{\rho_a}\iota_{\rho_b}H+2g_{[b}\iota_{\rho_{a]}}F-2f_{[b}\iota_{\rho_{a]}}\Omega+g_{[b}\dd f_{a]}+f_{[b}\dd g_{a]}\,,\nn\\
  \label{constraint2b}
  \eea 
  where in the second line we used the Cartan relation 
  \be 
  {\cal L}=\iota \circ \dd +\dd \circ \iota\,,
  \ee 
 and in the last line we also used the constraint \eqref{constraint1}, which may be written explicitly as 
 \be \label{constraint1b}
 \frac 12 \left(\iota_{\r_a}\theta_b+\iota_{\rho_b}\theta_a+f_ag_b+f_bg_a\right)=0\,.
 \ee 
 
 In order to provide a geometric interpretation of these results, let us recall the definitions of twisted contact Courant algebroids and their Dirac structures. Contact Courant algebroids twisted by a 3-form $H$ and two 2-forms $F$ and $\Omega$ are defined in Ref. \cite{BouwknegtLectures}{\footnote{The version in terms of non-skewsymmetric bracket is found in \cite{Wright}.}} as follows. Consider the vector bundle $E=TM\oplus \R\oplus T^{\ast}M\oplus \R$, whose sections are $\X=(X,f,\eta,g)$, where $X$ is a vector field, $\eta$ an 1-form and $f,g$ are functions. Then the twisted contact Courant algebroid is given by the data $(E,[\cdot,\cdot]_E,\langle\cdot,\cdot\rangle, a:E\to TM)$ of the above vector bundle, 
 a skewsymmetric bracket, a nondegenerate symmetric bilinear form and an anchor map, given as
 \bea 
 [\X_1,\X_2]_{E}&=&\left(	[X_1,X_2], X_1(f_2)-X_2(f_1)+\iota_{X_1}\iota_{X_2}F,\right. \nn\\ 
 && \left. \,\,\, {\cal L}_{X_{1}}\eta_{2}-{\cal L}_{X_2}\eta_1+g_{2}\iota_{X_{1}}F-g_1\iota_{X_2}F-\sfrac 12 \dd (\iota_{X_1}\eta_2-\iota_{X_2}\eta_1)\right.\label{twistedcontactbracket} \nn\\ 
 && \left. \,\,\, +\sfrac 12(g_2\dd f_1-g_1\dd f_2-f_1\dd g_2+f_2\dd g_1)-\iota_{X_1}\iota_{X_2}H
-f_2\iota_{X_1}\Omega+f_1\iota_{X_2}\Omega,\right. \nn\\ 
&&\left. \,\,\, X_1(g_2)-X_2(g_1)-\iota_{X_1}\iota_{X_2}\Omega\right)\,,\\ 
\langle\X_1,\X_2\rangle&=&\frac 12 \left(\iota_{X_1}\eta_2+\iota_{X_2}\eta_1+f_1g_2+f_2g_1\right)\,,\\ 
a(\X)&=&X\,,  \eea  
 with $\dd F=0=\dd\Omega$ and $\dd H=-F\w\Omega$. (Note that in comparison to \cite{BouwknegtLectures} we have a different sign convention for $H$ and $\Omega$.) It is now clear what the constraints we derived above mean in this geometric setting. First, the 1-forms $\theta_a$ and the functions $g_a$ can be associated to the following maps 
 \bea 
&&\theta: {\cal E} \to T^{\ast}M  \qquad \qquad \qquad g: {\cal E} \to \R \nn\\ 
&& \quad\, e_a\mapsto \theta(e_a):=\theta_a \qquad \qquad \,\,\, e_a \mapsto g(e_a):=g_a\,. 
 \eea 
 Together with $\rho$ and $f$, one then obtains a map 
 \bea 
&&\hat{\rho}:= \rho\oplus f \oplus\theta\oplus g:  {\cal E}\to E \nn\\ 
&&\qquad \qquad\,\,\,\, \qquad \qquad  e_a\mapsto \hat{\r}(e_a):=\rho_a+f_a+\theta_a+g_a:=\hat{\r}_a\,.
 \eea 
Then, combining \eqref{rhobracket} and \eqref{fbracket} with the two constraints written in the form \eqref{constraint3b} and \eqref{constraint2b}, we directly obtain 
\be 
[\hat{\r}_a,\hat{\r}_b]_{E}=C_{ab}^{c}\hat{\r}_c\,,
\ee 
in other words the generalised sections $\hat{\rho}$ as defined above are closed under the bracket of the twisted contact Courant algebroid. In addition, the remaining constraint \eqref{constraint1b} states that 
\be 
\langle\hat{\rho}_{a},\hat{\rho}_b\rangle =0\,.
\ee 
This means that the generalized sections $\hat{\rho}$ are constrained in a subbundle of $E$, which is isotropic and involutive with respect to the twisted Courant bracket \eqref{twistedcontactbracket}. 
Subbundles of the standard Courant algebroid on the vector bundle $TM\oplus T^{\ast}M$ with the above properties where first encountered in the context of constrained Hamiltonian systems via an analysis based on the Dirac bracket and are therefore called Dirac structures \cite{dirac}. Dirac structures interpolate between presymplectic and Poisson ones and they give rise to topological sigma models in two dimensions which were dubbed Dirac sigma models \cite{Kotov:2004wz}. Here we encounter the analogous structure for the contact Courant algebroid on $TM\oplus \R\oplus T^{\ast}M\oplus \R$, with the respective contact Dirac structures as constrained subbundles. 

From an alternative point of view, first proposed in \cite{Chatzistavrakidis:2017tpk} for the standard case, the gauging of the original sigma model can be interpreted as a constraining of a different action functional, which is essentially the dimensional reduction of the universal action functional described in \cite{Chatzistavrakidis:2017tpk}. The starting point for this interpretation is the introduction of two auxiliary 1-forms $V^{i}$ and $W_{i}$ taking values in $T^{\ast}M$ and $TM$ respectively, and two auxiliary functions $v$ and $w$. Then we can write the action functional 
\bea 
{\cal S}&=&-\int_{\S_2}\frac 12 G_{ij}\mathrm{D}X^{i}\w\ast\mathrm{D}X^{j}+
\frac 12 G_{\Phi\Phi}\,\l\w\ast\l -\int_{\S_3}H+\Omega\w\Lambda \nn\\[4pt] && - \int_{\S_2}W_i\w\left(\dd X^{i}-\frac 12 V^{i}\right)-\int_{S_{2}}w\w\left[\dd \Phi-\frac 12 v+\mf{a}_i\left(\dd X^{i}-\frac 12 V^{i}\right)\right]\,,
\eea 
where $\mathrm{D}X^{i}=\dd X^{i}-V^{i}$ and $\lambda=\dd\Phi-v+\mf{a}_i(\dd X^{i}- V^{i})$. 
This action functional has the property that when the auxiliary fields are unconstrained it is equivalent to the ungauged action functional $S_0$, whereas when the auxiliary fields are appropriately constrained it yields the gauged versions of the model corresponding to the action functional $S$. 
Regarding the first part of this statement, the field equations for the four auxiliary fields read
\bea 
&& V^{i}=2\dd X^{i}\,,\quad v+\mf{a}_iV^{i}=2(\dd\Phi+\mf{a}_{i}\dd X^{i})\,, \quad w=
2 G_{\Phi\Phi}\ast\l\,,\nn\\ 
&& W_i+\mf{a}_iw=2G_{ij}\ast\mathrm{D}X^{j}+\mf{a}_iG_{\Phi\Phi}\ast\l\,.
\eea 
Substituting them in $\cal S$, one directly obtains the ungauged action $S_0$. On the other hand, 
if the auxiliary fields are constrained to live on a Dirac structure of the contact Courant algebroid on $TM\oplus \R\oplus \R\oplus T^{\ast}M$, then ${\cal S}$ yields the gauged action $S$ upon the identifications 
\bea 
V^{i}=\rho^{i}_aA^{a}\,,\quad v=\widetilde{f}_aA^{a}\,,\quad W_{i}=\theta_{ai}A^{a}\,,\quad w=g_{a}A^{a}\,.
\eea 
This identification only works if $\gamma_{ab}=\iota_{\rho_{a}}\theta_b+f_ag_b$, which is precisely the relation obtained from the consistency of the gauging procedure. Thus we see that, in analogy to \cite{Chatzistavrakidis:2017tpk}, one may think of the gauging not as an extension of the given action $S_0$ by additional gauge fields but as a restriction of the equivalent action ${\cal S}$ to a constrained set of fields.

\subsection{The non-Abelian case and nonexact Courant algebroids}
\label{sec24}

Let us now examine the meaning of the constraints arising from the gauged action functional in the non-Abelian case. First of all,  condition \eqref{condition3} implies that 
\be \label{fa}
f_a^{\a}=\iota_{\rho_a}\mf a^{\a}\,. 
\ee 
Secondly, the fact that $\dd F^{\a}=-\dd \Omega^{\a}\ne 0$ prompts us to identify{\footnote{More precisely, at this stage the two 2-forms differ by an exact 2-form. The properties of Courant algebroids with nonclosed 3-form \cite{RCA} provide an a posteriori justification for the proportionality of the 2-forms $\Omega$ and $F$.}} 
\be 
\Omega^{\a}=-F^{\a}\,,\label{FOmega}
\ee 
therefore the first of \eqref{dtwists} becomes 
\be 
\dd \mc H=k_{\a\b}\, F^{\a}\w F^{\b}\,.
\ee 
Then the first constraint is written explicitly as 
\be 
\label{constraint1c}
\frac 12 \left(\iota_{\r_a}\theta_b+\iota_{\rho_b}\theta_a+k_{\a\b}f_a^{\a}g_b^{\b}+k_{\a\b}f_b^{\a}g_a^{\b}\right)=0\,.
\ee 
Furthermore, similar manipulations to the Abelian case lead to the following form of constraints 
\eqref{constraint2} and \eqref{constraint3} respectively, 
\bea 
2\iota_{\rho_{[a}} \dd g_{b]}^{\a}-\iota_{\rho_a}\iota_{\rho_b}\Omega^{\a}-K^{\a}{}_{\b\g}f_a^{\b}f_b^{\g}=C_{ab}^{c}g_c^{\a}\label{constraint3c}
\eea  
and 
\bea 
C_{ab}^{c}\theta_c&=&2{\cal L}_{\r_{[a}}\theta_{b]}-\dd\iota_{\rho_{[a}}\theta_{b]}-\iota_{\rho_a}\iota_{\rho_b}{\cal H} +2\left(g_{[b}^{\a}+f_{[b}^{\a}\right)\iota_{\rho_{a]}}F^{\a} 
 +g_{[b}^{\a}\nabla f_{a]\a}+f_{[b}^{\a}\nabla g_{a]\a}
\,,\nn\\ \label{constraint2c}
\eea 
where we have defined 
\be 
(\nabla g_{a})^{\a}:=\dd g_{a}^{\a}+K^{\a}{}_{\b\g}g_a^{\b}\mf a^{\g}\,,\label{connection}
\ee
and similarly for $f_{a}^{\a}$. In addition to these constraints, condition \eqref{fa} implies that:
\bea
\nonumber C^c_{ab}f_c^\alpha&=& C^c_{ab}\iota_{\rho_c}a^\alpha\overset{\eqref{rhobracket}}=\iota_{[\rho_a,\rho_b]}a^\alpha=\mathcal{L}_{\rho_a}\iota_{\rho_b}a^\alpha-\iota_{\rho_b}\mathcal{L}_{\rho_a}a^\alpha\\
\nonumber &=&\iota_{\rho_a}\dd\iota_{\rho_b}a^\alpha-\iota_{\rho_a}\dd\iota_{\rho_b}a^\alpha-\iota_{\rho_b}\iota_{\rho_a}\dd a^\alpha\\
\nonumber &=&2\iota_{\rho_{[a}}\dd f_{b]}^\alpha-\iota_{\rho_b}\iota_{\rho_a}F^\alpha+\tensor{K}{^\alpha_{\beta\gamma}}f_a^\beta f_b^\gamma\\
&=&2\iota_{\rho_{[a}}\left(\nabla f_{b]}\right)^\alpha-\iota_{\rho_b}\iota_{\rho_a}F^\alpha-\tensor{K}{^\alpha_{\beta\gamma}}f_a^\beta f_b^\gamma\,,\label{condition3c}
\eea
which is a non-Abelian version of the assumption \eqref{fbracket} that we have made in the beginning. This means that, unlike the Abelian case, in the non-Abelian case this follows directly from the conditions on the background fields.

In a similar spirit to the Abelian case, our goal now is to identify the resulting constraints in terms of a bracket and bilinear in an appropriate Courant algebroid. 
Note that exactly because of the additional structure $\mc G$, this  cannot be an exact Courant algebroid. Recall that a Courant Algebroid is called exact if its vector bundle $E$ fits into the exact sequence 
\be 
0 \, \to \, T^{\ast}M \, \overset{\rho^{\ast}}\to  \,E \, \overset{\rho}\to \, TM \, \to 0~, 
\ee 
where $\rho$ is the anchor map. Exact CAs are classified by the \v{S}evera class, a degree-3 class in the de Rham cohomology of the smooth manifold $M$ \cite{Severa:2017oew}. 

To understand the geometric meaning of the above result, we employ the definition of a regular Courant algebroid found in Ref. \cite{RCA}.  
First recall that the image of the anchor map $\rho$ generates a foliation of $M$. Let us denote the corresponding distribution as
\be 
{\cal F}:=\rho(E)\,.
\ee
A foliation is called regular if its leaves are of equal rank, otherwise it is called singular. Most foliations are clearly singular, however, if we concentrate on regular ones the corresponding Courant algebroid is called regular \cite{RCA}. 
In general, the anchor has a nontrivial kernel, due to the Courant algebroid property
\be 
\rho\circ{\cal D}f=0\,, 
\ee  for any smooth function $f$, where   
  the derivation ${\cal D}$ is defined through 
  \be \label{Ddef}
  \langle e,{\cal D}f\rangle =\sfrac 12 \rho(e)f\,,
  \ee 
  for every section $e$ of $E$ \cite{Liu:1995lsa}.
Moreover,  one can define the complement of the kernel, $(\text{ker} \, \rho)^{\perp}$, with respect to the bilinear form on $E$. A general statement is that the kernel of the anchor map is coisotropic, 
which means that
\be 
(\text{ker}\,\rho)^{\perp} \, \subset \, \text{ker}\,\rho~.
\ee
This is simple to prove, since by definition 
\be 
(\text{ker}\,\rho)^{\perp}=\{e\in E \, | \, \langle e,e'\rangle=0\,, \, \forall e'\in E \,\, \text{such that}\,\,  \rho(e')=0\}~.
\ee
From \eqref{Ddef} we directly see that every $e$ in the complement of the kernel necessarily belongs to the kernel itself.

For every regular Courant algebroid, the kernel and its complement are smooth subbundles of $E$ 
and the quotients of $E$ by them are Lie algebroids. In particular, there is a canonical isomorphism 
\be 
E/\text{ker}\,\rho \simeq {\cal F}~.
\ee 
On the other hand, the quotient $E/(\text{ker}\,\rho)^{\perp}:={\cal A}_{E}$ is also a Lie algebroid---it is called the ample Lie algebroid  of $E$ in \cite{RCA}. 
In addition, the quotient 
\be 
{\cal G}:=\text{ker}\,\rho/(\text{ker}\,\rho)^{\perp}
\ee 
is defined.
Obviously it is trivial when the complement of the kernel is all the kernel of the anchor. This is the case, for instance, if we consider the standard (and regular) Courant algebroid over $M$ with anchor being the projection to the tangent bundle $TM$; then the kernel is $T^{\ast}M$, its complement is also the same and ${\cal G}$ is trivial. On the other hand, for the contact Courant algebroid that we encountered in the previous section, the kernel of $\rho$  and its complement are 
\be 
\text{ker}\,\r=\R\oplus T^{\ast}M\oplus\R\,, \quad (\text{ker}\,\r)^{\perp}=T^{\ast}M\,,
\ee 
and therefore ${\mc G}$ is nontrivial, being $\R\oplus\R$. 
In general, using the projection $\pi: \text{ker}\,\rho \to {\cal G}$, one can turn ${\cal G}$ into a bundle of quadratic Lie algebras, i.e. a vector bundle with every fiber being a quadratic Lie algebra (with a bracket determined by the bracket on $E$) in the sense that it admits a nondegenerate, ad-invariant inner product on its vector space \cite{RCA}. For completeness, recall that every bundle of Lie algebras on a smooth manifold defines a Lie algebroid with zero anchor. 

In the case of the contact Courant algebroid, the vector bundle is thus identified with $TM\oplus {\mc G}\oplus T^{\ast}M$. It turns out that this holds more generally for regular Courant algebroids. Specifically, \cite{RCA} define an isomorphism $\Psi$ of vector bundles---which is called a dissection of $E$---as  
\be 
\Psi\,: \, {\cal F}^{\ast}\oplus {\cal G}\oplus {\cal F} \, \to E~, 
\ee 
such that 
\be \label{psi}
\langle \Psi(\eta_1+s_1+X_1), \Psi(\eta_2+s_2+X_2)\rangle=\sfrac 12 \big(X_1(\eta_2)+X_2(\eta_1)\big)+\langle s_1,s_2\rangle^{{\cal G}}~.
\ee
This isomorphism is useful in the case of nonexact Courant algebroids, since it offers an alternative definition which to some extent is better fitted to the structure. Instead of the quadruple $(E,[\cdot,\cdot]_E,\langle\cdot,\cdot\rangle,\rho)$ of a vector bundle, a skew-symmetric binary operation, a symmetric bilinear form and an anchor, satisfying the properties laid out for example in \cite{Liu:1995lsa}, one can consider a quintuple $({\cal F},{\cal G},\nabla,F,{\cal H})$. It comprises the distribution ${\cal F}$, the bundle of quadratic Lie algebras ${\cal G}$ and three canonical maps.  A connection 
	\bea 
	 \nabla \, :\, \G({\cal F})\otimes \G({\cal G}) \to \G({\cal G}) 
	\eea 
	satisfying linearity and Leibniz rule
	\bea 
	\nabla_{fX}s&=&f\nabla_{X}s\,,
	\\ 
	\nabla_{X}(fs)&=&f\nabla_{X}s+(X(f))s\,,
	\eea
a $C^{\infty}$-bilinear map (analogous to a ${\mc G}$-valued curvature 2-form)
	\bea 
	 F\,:\, \G(\w^2{\cal F})\to \G({\cal G}) 
	\eea
	and a $C^{\infty}$-bilinear map (analogous to a 3-form)
	\bea 
	 {\cal H}\,:\, \otimes^3 \G({\cal F}) \to C^{\infty}(M)\,. 
	\eea
	This structure is a Courant algebroid if a number of additional properties are satisfied (see \cite{RCA} for a detailed exposition of the corresponding identities):
	\begin{subequations}\label{Bconditions}
	\bea 
\dd{\cal H}&=&\langle F\w F\rangle\,,\label{Bcondition1}\\
\mathcal{L}_{\rho_a}\langle s_b,s_c\rangle^\mathcal{G}&=&\langle\nabla_{\rho_a}s_b,s_c\rangle^\mathcal{G}+\langle s_b,\nabla_{\rho_a}s_c\rangle^\mathcal{G}\,,\label{Bcondition2}\\
\mathcal{L}_{\rho_a}[s_b,s_c]^\mathcal{G}&=&[\nabla_{\rho_a}s_b,s_c]^\mathcal{G}+[s_b,\nabla_{\rho_a}s_c]^\mathcal{G}\,,\label{Bcondition3}\\
\nabla_{\rho_a}F(\rho_b,\rho_c)-F([\rho_a,\rho_b],\rho_c)+c.p.&=&0\,,\label{Bcondition4}\\
2\nabla_{\rho_{[a}}\nabla_{\rho_{b]}}s_c-\nabla_{[\rho_a,\rho_b]}s_c&=&[F(\rho_a,\rho_b),s_c]^\mathcal{G}\,,\label{Bcondition5}
	\eea 
	\end{subequations}
	where the 4-form on the right-hand side is given as 
	\be 
	\langle F,F \rangle (X_1,X_2,X_3,X_4):=\frac 14 \sum_{\s} \text{sgn}(\s)\langle F(X_{\s(1)},X_{\s(2)}),F(X_{\s(3)},X_{\s(4)}\rangle^{\cal G}\,,
	\ee 
and $\s$ runs over all permutations of four objects. In this setting, the standard (nonexact) Courant algebroid is equipped with the following Courant bracket,
\bea 
[\X_1,\X_2]&=&\left([X_1,X_2],\nabla_{X_1}s_2-\nabla_{X_2}s_1+ [s_1,s_2]^{\cal G}+\iota_{X_1}\iota_{X_2}F,\right. \nn\\ 
&& \left.\,\,\, {\cal L}_{X_1}\eta_2-{\cal L}_{X_2}\eta_1-\sfrac 12 \dd\left(\iota_{X_1}\eta_2-\iota_{X_2}\eta_1\right)-\iota_{X_1}\iota_{X_2}{\cal H}\right.\nn\\ 
&& \left.\,\,\, + \left(\langle \nabla s_1,s_2\rangle^{\cal G}-\langle \nabla s_2,s_1\rangle^{\cal G}\right) +2\langle s_1,\iota_{X_2}F\rangle^{\cal G} -2\langle s_2,\iota_{X_1}F\rangle^{\cal G}\right)\,,
\label{CSXbracket}\eea 
for sections $\X=(X,s,\eta)\in  {\cal F}\oplus {\cal G}\oplus {\cal F}^{\ast}$.

We are now ready to provide an interpretation of the gauging constraints in terms of the structures reviewed here. To this end we consider ${\cal G}={\cal G}_{L}\oplus {\cal G}_{R}$ for the bundle of quadratic Lie algebras, and decompose its sections accordingly as $s=(f,g)$ with $f\in {\cal G}_{L}$ and $g\in {\cal G}_R$.  For the inner product in ${\cal G}$ we identify 
\be 
\langle s_a,s_b\rangle^{\cal G}=\frac 12 k_{\a\b}(f_{a}^{\a}g_{b}^{\b}+g_a^{\a}f_b^{\b})\,.
\ee 
Then the first constraint \eqref{constraint1c} takes the general form 
\be 
\langle \Psi(\hat\r),\Psi(\hat\r') \rangle =0
\ee 
in terms of the dissection $\Psi$, where $\hat\r,\hat\r'\in \G(E)$\,. 
Furthermore, comparing the bracket \eqref{CSXbracket} and the gauging constraints \eqref{constraint3c} and  \eqref{constraint2c}, the above identification directly leads to the closure of the bracket, namely 
\be 
[\hat\r_a,\hat\r_b]=C_{ab}^{c}\hat\r_c\,,
\ee 
with the Lie bracket in ${\cal G}$ being 
\be 
[(f_a,g_a),(f_b,g_b)]^{\a}=K^{\a}{}_{\b\g}(f_a^{\b}f_b^{\g},f_a^{\b}f_b^{\g}+f_a^{\b}g_b^{\g}-f_b^{\b}g_a^{\g})\,.
\ee 
All properties \eqref{Bconditions} are satisfied with this identification, as they correspond to the conditions \eqref{conditions} on the background fields and the choice of connection \eqref{connection}. This directly generalizes the corresponding result of the Abelian case. The fields of the gauged theory are constrained on Dirac structures of the nonexact Courant algebroid over $TM\oplus {\cal G}\oplus T^{\ast}M$.

Before we close this section, it is worth comparing in more general terms the Abelian and non-Abelian cases. It should be clear that the Abelian case is not fully contained in the non-Abelian one. One difference, as already mentioned, is the absence of condition \eqref{condition3} in the Abelian case. As a result, condition \eqref{condition3c} is absent in the Abelian case. Furthermore, the fields $F$ and $\Omega$ are independent in the Abelian case, while in the non-Abelian they are related through \eqref{FOmega}. Finally, it would seem that the Courant bracket obtained in the Abelian version is different from the non-Abelian one. However, the way that the Courant bracket is defined here does include the Abelian case as well. It turns out that there is more freedom in satisfying properties \eqref{Bconditions} if $F$ is closed since in that case it is possible for $F$ to act independently on $\mathcal{G}_L$ and $\mathcal{G}_R$. Specifically, we take
\be
F((f_a,g_a),(f_b,g_b))=(F(f_a,f_b),-\Omega(g_a,g_b))\,.
\ee
In the non-Abelian case this is not possible since the bracket would not satisfy all of the necessary properties.

\subsection{Examples}

Let us now discuss a couple of simple examples of gauging along vector fields $\rho_a$ in the presence of a nonclosed 3-form $H$. Both examples refer to the Abelian case of subsection \ref{sec23} and they differ in that $\rho_a$ form an Abelian algebra in the first and a non-Abelian one in the second. 

In both examples, we take as $M$ a real, Euclidean 4-manifold with coordinates $x^{i}=(x,y,z,w)$---and we denote the corresponding pull-backs $X^{i}$ with the same lower case letters. The additional direction (of the circle bundle) is denoted by $\Phi$ as before. Furthermore, we consider the following 3-form and 2-forms 
\bea \label{examplesH}
H=x\,\dd y\w\dd z\w\dd w\,, \quad \Omega=\dd y\w\dd z\,, \quad F=-\dd x\w\dd w\,.
\eea 
Evidently, both $\Omega$ and $F$ are closed 2-forms, whereas the 3-form $H$ is not closed and it satisfies 
$\dd H=-F\w\Omega$. We should also specify the manifold, in particular its metric $G$, which we do separately in each case. 

First, we consider a single vector field $\rho_x$, specifically 
\be 
\rho_x = y\partial_z-z\partial_y\,.
\ee 
This vector field generates rotations and any manifold $M$ with metric $G$ which is invariant under this rotation, namely for which $\rho_x$ is a Killing vector field, is admissible. Here we consider the flat metric for simplicity, 
\be 
\dd s^2 = \dd x^2+\dd y^2+\dd z^2+\dd w^2\,.
\ee 
Our goal now is to specify $f_x$, $g_x$ and $\theta_x$ such that all consistency conditions \eqref{conditions} and constraints \eqref{constraints} for gauging along the vector field $\rho_x$ are satisfied---in the present Abelian case, for $K_{\a\b\g}=0$ and $\a$ taking only one value. First, from \eqref{condition2}, we can determine the function $g_x$, at least up to exact terms. We find 
\be 
g_x=-\frac {y^2+z^2}{2}\,.
\ee 
It may be directly checked that the constraint \eqref{constraint3} is then identically satisfied. 
Next, condition \eqref{condition1} is consistent only if $\dd f_x=0$, as can be checked by acting on it with the exterior derivative $\dd$. Thus, $f_x$ should be a real constant. Then, one can determine the 1-form $\theta_x$ up to exact terms, 
\be 
\theta_x=-\frac x2 \left(y^2+z^2\right)\dd w+\frac f2 \left(y\dd z-z \dd y\right)\,.
\ee  
This makes sure that the constraints \eqref{constraint1} and \eqref{constraint2} are identically satisfied, therefore all the necessary and sufficient conditions for consistent gauging hold. The gauged action functional, containing the single gauge field $A_x$, reads explicitly as 
\bea 
S&=&- \int_{\S_2} ||\dd x||^2+||\dd y+zA_x)||^2+||\dd z-yA_x||^2+||\dd w||^2\nn\\ 
&&-\int_{\S_2}\frac f2 A_x\w\left(y\dd z-z\dd y\right)+\frac x2 (y^2+z^2)A_x\w \dd w -\frac 12 (y^2+z^2)A_x\w \Lambda\nn\\ &&-\int_{\S_3}x\,\dd y\w\dd z\w\dd w+\dd y\w \dd z\w \Lambda\,,
\label{ex1}\eea
where we denote the inner product $||\omega||^2=\frac 12 \omega\w\ast\omega$ and the fiber 1-form is 
$\Lambda=\dd \Phi-x\dd w$. The infinitesimal gauge transformations are 
\be 
\d y=-z\epsilon\,, \quad \d z=y\epsilon\,,\quad \d\Phi=f\epsilon\,,\quad  \d A_x=\dd \epsilon\,,
\ee 
where $\epsilon$ is the single gauge parameter of the model.
It is worth emphasizing that from the point of view of the higher-dimensional geometry of the target space   $M\times S^{1}$, the 3-form supported on $\S_3$ is closed; indeed, it can be written as $\dd y\w\dd z\w\dd \Phi$, which is then the Wess-Zumino term for the action $S$. However, from a dimensional reduction perspective, lower dimensional fields are not identified using $\dd\Phi$ as an expansion 1-form but using instead the 1-form $\Lambda=\dd\Phi+\mf{a}$, where presently $\mf{a}=-x\dd w$ is the ``Kaluza-Klein'' gauge field. This is precisely what we did above by identifying $H$ and $\Omega$ as in \eqref{examplesH}. Indeed, then the full 3-form in the higher-dimensional theory is closed, whereas the 3-form of the dimensionally reduced theory is not and it is given as above. It is precisely this Kaluza-Klein perspective that prompted us to write the Wess-Zumino term in \eqref{ex1} as such.

As a second example, we would like to consider a set of non-Abelian vector fields $\rho_a$. To this end, we consider the same background fields in the topological sector as in the first example, and the vector fields 
\be 
\rho_x=\partial_x\,, \qquad \rho_y=\partial_y+z\partial_x\,,\qquad \rho_z=\partial_z\,.
\ee 
We observe that there is one nonvanishing commutator, namely 
\be 
[\rho_{y},\rho_{z}]=-\rho_x\,,
\ee 
therefore one nonvanishing structure constant $C_{zy}^{x}=1$. This is nothing but a three-dimensional nilpotent Lie algebra. The background metric should then be invariant under this symmetry, and this is indeed the case for the following metric, 
\be 
\dd s^2= (\dd x-z\dd y)^2+\dd y^2 +\dd z^2+\dd w^2\,.
\ee 
Given the above background data, one may readily check that all conditions \eqref{conditions} and constraints \eqref{constraints} are satisfied for the following 
\bea 
&& f_x=0\,, \qquad \,\,f_y=-1=f_z\,,\\
&& g_x=1\,,\qquad \, \,g_{y}=z\,, \qquad g_{z}=-y\,,\\
&& \theta_x=x\dd w\,, \quad \theta_{y}=\dd x-y\dd z+xz\dd w\,,\quad \theta_{z}=\dd x-y\dd z-xy\dd w\,.
\eea   
This leads to a gauged action functional of the form,
\bea 
S&=&-\int_{\S_2}||\dd x-z\dd y-A_x||^2+||\dd y-A_y||^2+||\dd z-A_z||^2+||\dd w||^2\nn\\ 
&& -\int_{\S_2} x A_x\w\dd w+A_y\w (\dd x-y\dd z+x\,z\dd w)+A_{z}\w(\dd x-y\dd z-x\,y\dd w)\nn\\ 
&&
- \int_{\S_2} \left(A_x+zA_y-yA_z\right)\w \Lambda +A_x\w (A_y+A_z)- (y+z)A_y\w A_z\nn\\ 
&&-\int_{\S_3}x\,\dd y\w\dd z\w\dd w+\dd y\w \dd z\w \Lambda\,,
\eea 
where similar comments to the ones below \eqref{ex1} apply here. 

\section{Jacobi structures, Jacobi sigma models \& twists}
\label{sec3}

Contact structures, such as the ones encountered in section \ref{sec23}, may be viewed as special cases of Jacobi structures for odd dimensional manifolds. Our purpose in this section is to investigate which sigma models are associated to Jacobi structures and whether such theories can be twisted in a sense to be explained below.\footnote{Soon after our paper was completed, the paper \cite{jsm} on Jacobi sigma models appeared, with some overlapping results with our sections \ref{sec31} and \ref{sec32}.} In essence, the analysis and results of the present section are independent from section \ref{sec2}. However, they can be viewed as a natural generalisation in the direction of allowing the background fields to depend on the additional spacetime direction. This is in contrast to trivial dimensional reduction and we comment on this in the following. 

\subsection{Jacobi structures}\label{sec31}

Let us first recall some basic facts about Jacobi structures. A Jacobi manifold $(M,\Pi,V)$ is a smooth manifold $M$ endowed with a pair $(\Pi,V)$ of a bivector field $\Pi\in\G(\bigwedge^2TM)$ and a vector field $V\in\G(TM)$, which satisfy the conditions \cite{Jacobi}
\be \label{Jacobiconditions}
[\Pi,\Pi]_{\text{S}}=2V\w\Pi\quad \text{and}\quad [\Pi,V]_{\text{S}}=0~,
\ee 
where $[\cdot,\cdot]_{\text{S}}:\bigwedge^pTM\otimes\bigwedge^qTM\to \bigwedge^{p+q-1}TM$ is the Schouten-Nijenhuis bracket of multivector fields.
Jacobi manifolds are a simultaneous generalization of Poisson manifolds (obtained when the vector field vanishes), manifolds with a vector field (when the bivector vanishes), and contact manifolds (when $V\w \Pi^n\ne 0$ and $\text{dim}\,M=2n+1$; $V$ is then the Reeb vector field). The Lie bracket on the space of functions that generalizes the Poisson bracket is then given as 
\be 
\{f,g\}_{_{\text{J.B.}}}=\Pi(\dd f,\dd g)+f\, \iota_{V}\dd g-g\,\iota_{V}\dd f\,.
\ee 
Unlike the Poisson bracket, the Jacobi bracket does not satisfy the Leibniz rule for derivations. By direct computation one finds instead 
\be 
\{f,hg\}_{_\text{J.B.}}=\{f,h\}_{_\text{J.B.}}g+h\{f,g\}_{_\text{J.B.}}+hg\,\iota_{V}\dd f\,.
\ee 
However, the Jacobi bracket is obviously antisymmetric and satisfies the Jacobi identity. The latter is equivalent to  conditions \eqref{Jacobiconditions}. The space of smooth functions on a Jacobi manifold equipped with a Jacobi bracket is then a local Lie algebra as suggested by Kirillov \cite{kirillov}. 

As mentioned in the Introduction, multiplication of a Poisson structure by a function results in a Jacobi structure. Indeed, assuming that a bivector $\mc{P}$ on $M$ is Poisson and thus it satisfies 
\be \label{poisson}
[\mc P,\mc P]_{\text{S}}=0\,,
\ee 
we may use it to construct the bivector 
\be 
\Pi=f\mc P\,,\quad  f:M\to \R\,.
\ee 
However, clearly the new bivector does not have a vanishing Schouten-Nijenhuis bracket with itself; it is straightforward to calculate
\be 
[\Pi,\Pi]_{\text{S}}=f\partial_if\mc{P}^{ij}\mc{P}^{kl}\partial_j\w\partial_k\w\partial_l\equiv 2V\w\Pi\,,
\ee 
with the identification
\be \label{VPi}
V=\partial_if\mc{P}^{ij}\partial_j=\mc {P}(\dd f)=f^{-1}\Pi(\dd f)\,,
\ee 
assuming that $f$ is invertible. This is a first relation between Poisson and Jacobi structures. 
However, there exists a second interesting relation between these two structures. Given the data $(\Pi,V)$ of a Jacobi structure on a smooth manifold $M$, one can construct a Poisson structure on the manifold $\widetilde{M}=M\times \R_{+}$ with bivector field
\be 
\widetilde{\mc P}:=e^{-\Phi}\left(\Pi+\frac{\partial}{\partial\Phi}\w V\right)\,,
\ee  
and $\Phi$ being the coordinate on $\R$.
Indeed, one can directly check that the Schouten-Nijenhuis bracket of $\widetilde{\mc P}$ with itself vanishes and thus it defines a Poisson bracket on $\widetilde{M}$. This is called Poissonization of the Jacobi structure. In physics parlance it corresponds to a dimensional oxidation, the procedure of obtaining a higher-dimensional theory from a lower-dimensional one, i.e. the inverse of dimensional reduction. 
We will find a field-theoretical incarnation of this feature in the following subsection.

\subsection{Jacobi Sigma Models} \label{sec32}

Our main goal here is to investigate the construction and  symmetries of nonlinear sigma models when the target space is associated to a Jacobi structure.  
In other words, we wish to show that there is a correspondence between Jacobi manifolds and a certain class of sigma models, henceforth called Jacobi Sigma Models. 

In order to proceed,
one may start with the Poisson sigma model \cite{SchallerStrobl,Ikeda}, whose action functional is simply 
\be \label{psm}
S_{\text{PSM}}[X,A]=\int_{\S_2}A_i\w\dd X^i+\sfrac 12 \mc P^{ij}(X)A_i\w A_j\,,
\ee 
where $A_i$ are 1-forms on the two-dimensional source space $\S_2$ and $\mc P^{ij}(X)=X^{\ast}\mc P^{ij}(x)$ are pull-back functions of the components of the Poisson bivector on the target space $M$ with coordinates $x^{i}$ by the pull-back function of the sigma model map $X:\S_2\to M$. Recall that the differential $\dd X^i$ reads explicitly as $\dd X^i=\partial_{\a}X^i\dd\s^{\a}$, where $\s^{\a}, \a=0,1$ are local coordinates on the two-dimensional worldsheet. The above action functional is invariant under the following set of infinitesimal gauge transformations, 
\bea 
\d X^i&=&\mc P^{ji}\e_{j}\,,\\
\d A_i&=&\dd\e_i+\partial_i\mc P^{jk}A_j\e_k\,,
\eea 
with $\S_2$-dependent scalar gauge parameter $\e_i$. Gauge invariance is guaranteed precisely by the property of $\mc P$ being Poisson, i.e. satisfying \eqref{poisson}. Note that this model may be obtained via gauging as well, precisely in the spirit of section \ref{sec2}, ignoring the metric term and the 3-form and 2-form twists. One should then consider the Lie algebroid $\cal E = T^{\ast}M$, with anchor being given in terms of the Poisson structure and bracket being the Koszul-Schouten one. Identifying $\rho_a^{i}$ with the components ${\cal P}^{ij}$ of the anchor and $\theta_{ai}$ with the constant $\d_i^{j}$, one indeed ends up with the Poisson sigma model after gauging. More details are found in \cite{Chatzistavrakidis:2016jci}. 

Motivated by the discussion in Section \ref{sec31}, we consider multiplying the Poisson structure with a function $f$, which we henceforth parametrize  as $f=e^{-\Phi}$ for later convenience. This will directly spoil the symmetries of the Poisson sigma model, and in order to restore gauge invariance one must add a BF-type term involving the function $f$ and a topological term based on the (pull-back of) the components of a vector field $V$.  In order to be explicit, we should note that eventually the function $f$, or equivalently $\Phi$, will depend on the local coordinates $\s^{\a}$ of the two-dimensional worldsheet. 
Here, in a similar spirit to section \ref{sec2}, we consider only an explicit such dependence, thus $\Phi=\Phi(\s)$ being a map from $\S_2$ to $\R$. 
Since we are still considering the sigma model map $X:\S_{2}\to M$ from the two-dimensional source manifold $\S_2$ to the $d$-dimensional target manifold $M$, one may also think in terms of the combination of the $d+1$ scalar fields to a map $\hat X=(X^i,\Phi):\S_2\to M\times \R$; this points to a higher-dimensional viewpoint where the target space is $(d+1)$-dimensional, as was the case in the concept of Poissonization discussed at the end of the previous subsection.

The action 
functional we consider for a Jacobi sigma model is 
\be \label{jsm}
S_{\text{JSM}}[X,\Phi,A,A_0]=\int_{\S_2}A_i\w\dd X^i+A_0\w\dd\Phi+\sfrac 12e^{-\Phi} \Pi^{ij}(X)A_i\w A_j+e^{-\Phi}V^i(X)A_0\w A_i~.
\ee
One may  view the combination of the worldsheet 1-forms $A_i$ and $A_0$ as taking values in $T^{\ast}M\oplus_{M}\R$. The background fields now are $\Pi^{ij}(X)$ and $V^i(X)$. Note though that the vector field $V$ is arbitrary and it is not given in terms of $\Pi$ as in \eqref{VPi}. 
In order to interpret the action functional \eqref{jsm} as a nonlinear gauge theory, we now consider the following infinitesimal gauge transformations,
\bea 
\d X^i&=&e^{-\Phi}\left(\Pi^{ji}\e_j+V^i\e_0\right)~,\\[4pt]
\d\Phi&=&-e^{-\Phi}V^i\e_i~,\\[4pt]
\d A_i&=&\dd\e_i+e^{-\Phi}\partial_i\Pi^{jk}A_j\e_k-e^{-\Phi}\partial_iV^j \left(A_j\e_0-A_0\e_j\right)\,,
\\[4pt]
\d A_0&=&\dd\e_0-e^{-\Phi}\Pi^{jk}A_j\e_k+e^{-\Phi}V^j\left(A_j\e_0-A_0\e_j\right)\,,
\eea
where $\epsilon_i$ and $\epsilon_0$ are two $\S_{2}$-dependent scalar gauge parameters and $\d:=\d_{(\epsilon_i,\epsilon_0)}$.  Our aim is to find under which conditions these gauge transformations leave the action functional \eqref{jsm} invariant. 
By straightforward calculations, the transformation of \eqref{jsm} is found to be  
\bea 
\d S_{\text{JSM}}&=&\int_{\S_2}e^{-2\Phi}\left(\e_iA_j\w A_k\left(\sfrac 12\Pi^{i{l}}\partial_l\Pi^{jk}+\Pi^{kl}\partial_l\Pi^{ij}+\sfrac 12 V^{i}\Pi^{jk}+V^k\Pi^{ij}\right)+\nn\right.\\
&& \qquad\qquad +\,\e_0A_i\w A_j\left(\sfrac 12 V^k\partial_k\Pi^{ij}+\Pi^{jk}\partial_kV^i\right)+\nn\\[4pt]
&&\left.\qquad\qquad +\,\e_iA_0\w A_j\left(\Pi^{ik}\partial_kV^j-\Pi^{jk}\partial_kV^i-V^k\partial_k\Pi^{ij}\right)\right)\,.
\eea
Using the local coordinate expression for the Schouten-Nijenhuis bracket for multivector fields, see e.g. \cite{Vaisman}, we infer that for $\Pi=\sfrac 12 \Pi^{ij}\partial_i\w\partial_j$ and $V=V^i\partial_i$,
\bea 
[\Pi,V]_{\text{S}}&=&\left(\sfrac 12 V^i\partial_i\Pi^{jk}+\Pi^{ij}\partial_iV^k\right)\partial_j\w\partial_k\,,\\[4pt]
{[}\Pi,\Pi]_{\text{S}}&=&\Pi^{li}\partial_l\Pi^{jk}\partial_i\w\partial_j\w\partial_k\,.
\eea
This immediately shows that the action functional is gauge invariant if and only if the bivector $\Pi$ and the vector $V$ constitute a Jacobi structure, i.e. 
\be 
\d S_{\text{JSM}}=0\quad \Leftrightarrow \quad [\Pi,V]_{\text{S}}=0 \quad \text{and} \quad [\Pi,\Pi]_{\text{S}}=2V\w\Pi\,,
\ee
which justifies the name of the model under consideration. Note that the target space in this case is $M\times \R$, with $M$ being a Jacobi manifold.

For completeness, one may proceed to find the equations of motion of the model and the structure of its gauge algebra. The field equations for the action functional \eqref{jsm} are 
\bea 
&&\frac{\d S_{\text{JSM}}}{\d A_i}\equiv DX^i:=\dd X^i+e^{-\Phi}\Pi^{ij}(X)A_j-e^{-\Phi}V^i(X)A_0=0\,,\label{eom1}\\[4pt]
&& \frac {\d S_{\text{JSM}}}{\d X^i}\equiv DA_i:=\dd A_i+\sfrac 12 e^{-\Phi}\partial_i\Pi^{jk}(X)A_j\w A_k+e^{-\Phi}\partial_iV^j(X)A_0\w A_j=0\,,\\[4pt]
&& \frac{\d S_{\text{JSM}}}{\d A_0}\equiv D\Phi:= \dd \Phi+e^{-\Phi}V^i(X)A_i=0\,,\label{eom3}\\[4pt]
&&\frac{\d S_{\text{JSM}}}{\d \Phi}\equiv DA_0:=\dd A_0-\sfrac 12 e^{-\Phi}\Pi^{ij}(X)A_i\w A_j-e^{-\Phi}V^i(X)A_0\w A_i=0\,.
\eea
One can check that the equations of motion indeed transform covariantly, in other words $D$ acting on the fields is a covariant worldsheet derivative.
For example, using the Jacobi structure conditions, we find
\bea
\nonumber\d\,D X^i &=&e^{-\Phi}\left(\epsilon_k\partial_j\Pi^{ki}+\epsilon_0\partial_j V^i\right)D X^j+e^{-\Phi}\left(\Pi^{ij}\epsilon_j-V^i\epsilon_0\right)D\Phi\,,\\[4pt]
\nonumber\d\,D\Phi &=& e^{-\Phi}\epsilon_iV^iD\Phi-e^{-\Phi}\epsilon_j\partial_i V^jD X^i\,.
\eea 

Much like the Poisson sigma model, the gauge algebra in the Jacobi case is open, in other words it only closes on-shell. The commutator algebra is found to be 
\bea 
[\d(\e),\d(\e')]X^i&=&\d(\e'')X^i \,, \\[4pt] {[}\d(\e),\d(\e')]\Phi&=& \d(\e'')\Phi\,,\\[4pt] {[}\d(\e),\d(\e')]A_i&=&\d(\e'')A_i+\e''_i D\Phi+e^{-\Phi}\left(\partial_iQ_j{}^{kl}\e_k\e'_l-\partial_i\widetilde{Q}_j{}^k(\e_k\e'_0-\e_0\e'_k)\right)DX^j\,,\\[4pt] 
{[}\d(\e),\d(\e')]A_0&=&\d(\e'')A_0-\e''_0D\Phi
-\e''_iDX^i\,,
\eea  
with the gauge parameters on the right hand side being 
\bea 
&&\e''_i:=e^{-\Phi}\left(Q_i{}^{jk}\e_j\e'_k-\widetilde{Q}_i{}^j(\e_j\e'_0-\e_0\e'_j)\right)\,,\\
&& \e''_0:=-e^{-\Phi}\left(\Pi^{ij}\e_i\e'_j-V^i(\e_i\e'_0-\e_0\e'_i)\right)\,,
\eea 
where we have defined the structure functions
\be 
Q_i{}^{jk}:=\partial_i\Pi^{jk} \quad \text{and} \quad \widetilde{Q}_i{}^j:=\partial_iV^j\,.
\ee 
Moreover, as is typical for gauge theories, there are also trivial gauge symmetries (with gauge transformations being proportional to the field equations, which is always possible).

Finally, one can now see that Poissonization applies directly to the two-dimensional sigma model presented here. Recall that we defined the map $\hat{X}=(\hat X^{I})=(X^{i},\Phi):\S_2\to M\times \R$, with $I=1,\dots,\text{dim}\,M+1$. Similarly, the worldsheet 1-forms are packaged accordingly as $A_{I}=(A_i,A_0)$. Then the action functional of the Jacobi sigma model may be rewritten as 
\be 
S_{\text{JSM}}[\hat X,A]=\int_{\S_2}A_I\w\dd \hat X^I+\sfrac 12\,\widetilde{ \mc P}^{ij}(\hat X)A_I\w A_J~,
\ee 
which is a Poisson sigma model with target space $\widetilde{M}$, distinct from the one defined by \eqref{psm}. 

\subsection{Examples}\label{sec33}

\subsubsection{Conformal symplectic manifolds}\label{sec331}

 A list of examples of Jacobi structures may be found for instance in Ref. \cite{marle}. One of the basic classes is locally conformal symplectic manifolds.
 Assume that $M$ is an even-dimensional manifold. It is a locally conformal symplectic manifold if it is endowed with a pair $(\Omega,\o)$ of a nondegenerate 2-form and a globally defined 1-form (called the Lee form) respectively, such that 
\be 
\dd\o=0\quad \text{and}\quad \dd\Omega+\o\w\Omega=0\,.
\ee
This structure can be obtained from a Jacobi structure as follows. One defines the vector field $V$ and the bivector field $\Pi$ to satisfy 
\be \label{lcjacobi}
i_{V}\Omega=-\o\quad \text{and}\quad i_{\Pi(\xi)}\Omega=-\xi\, 
\ee 
for every $\xi\in T^{\ast}M$. As such, $\omega$ and $\Pi$ satisfy 
\be \label{omegapi}
\Omega \circ \Pi=\text{Id} \quad \Rightarrow \quad \Omega_{ij}\Pi^{jk}=\d_{i}^{k}\,,  
\ee  
where we presented the component expression for clarity and because it will be useful in the following. Note that in the index-free equation \eqref{omegapi} we made an abuse of notation, in that $\Omega$ and $\Pi$ denote also the corresponding isomorphisms from $TM$ to $T^{\ast}M$ and vice versa. 

Let us now examine the corresponding Jacobi sigma model. We consider the equations of motion for $A_i$ and $A_0$, given by Eqs. \eqref{eom1} and \eqref{eom3} respectively. Since $A_i$ appears only algebraically, if we are able to solve for it we could substitute it back into the action. Since we have assumed nondegeneracy, this is presently possible. First, using \eqref{omegapi}, we may rewrite \eqref{eom1} as 
\be \label{ai0}
A_i=-e^{\Phi}\Omega_{ij}\dd X^j+\o_i A_0\,.
\ee 
where in the last term on the right-hand side we also used the first defining equation \eqref{lcjacobi}.  Turning then to the second field equation \eqref{eom3}, substituting \eqref{ai0} and using once more the defining conditions for the Jacobi structure, we readily obtain 
\be 
\o=-\,\dd\Phi\,.
\ee 
Plugging these equations back in the original action functional of the Jacobi sigma model, one obtains as a result a sigma model with locally conformal symplectic target given by 
\be 
S_{\text{LCSSM}}=\int_{\S_2}\sfrac 12 e^{\Phi}\Omega_{ij}(X)\dd X^i\w\dd X^j+2\,\o(X)\w A_0\,.
\ee 
Moreover, one may observe that although $\Omega$ is not a symplectic form, $\widetilde{\Omega}=e^{\Phi}\Omega$ is.  
On the other hand, setting $\omega$ to zero, one obtains the starting point of the topological A-model \cite{Witten:1988xj}.

\subsubsection{Contact manifolds}\label{sec332}

Another example of Jacobi structure is given by a contact manifold. Let $M$ be a manifold of dimension $2n+1$ equipped with a (contact) 1-form $\omega$ such that $\omega\wedge (\dd\omega)^n$ is nowhere vanishing. The Reeb vector field can then be defined by the conditions:
\begin{equation}
\iota_V\omega=1\quad\text{and}\quad\iota_V\dd\omega=0\,,
\end{equation}
while the bivector field of the Jacobi structure is defined through:
\begin{equation}
\iota_{\Pi(\xi)}\omega=0\quad\text{and}\quad\iota_{\Pi(\xi)}\dd\omega=-\xi+(\iota_V\xi)\omega\,,
\end{equation}
for all 1-forms $\xi$. These conditions can be written in component form:
\begin{eqnarray}
\Pi^{ij}\omega_j&=&0\,,\\
2\Pi^{jk}\partial_{[k}\omega_{i]}&=&-\delta^j_i+V^j\omega_i\,.
\end{eqnarray}

We can now examine the corresponding Jacobi sigma model. Multiplying the equation of motion for $A_i$, given by Eq. \eqref{eom1}, with $\omega_i$, we get an expression for $A_0$:
\begin{equation}
A_0=e^{\Phi}\omega_i\dd X^i\,.
\end{equation}
On the other hand, if we multiply the equation of motion for $A_i$ with $(\dd\omega)_{ik}$ and use the equation of motion for $A_0$, given by Eq. \eqref{eom3}, we get an expression for $A_i$:
\begin{equation}
A_i=e^{\Phi}(\dd\omega)_{ij}\dd X^j-e^{\Phi}\omega_i\dd\Phi\,.
\end{equation}
Plugging these equations back in the action of the Jacobi sigma model, and assuming a nontrivial boundary for $\S_2$, we obtain a 1-dimensional sigma model:
\begin{equation}
S_{\omega}=\int_{\Sigma_2}-\dd\left(e^{\Phi}\omega_i\dd X^i\right)=-\int_{\partial\Sigma_2}e^{\Phi}\omega_i\dd X^i\,.
\end{equation}
Thinking in terms of a nontrivial path $\gamma=\partial\S_2$, and neglecting the additional scalar field $\Phi$, the action functional 
\bea 
S_{\omega}: C^{\infty}(S^{1},M) &\to& \R \nn\\
\gamma &\mapsto& \int_{\g}\omega\,,
\eea
determines the closed trajectories of the Reeb vector field, and therefore the Reeb dynamics \cite{bourgeois}.

\subsection{Twisting the Jacobi Sigma Model}
\label{sec34}

In Ref. \cite{Klimcik:2001vg}, the authors considered a generalization of the Poisson sigma model in the spirit of WZW models by assuming that the target space is equipped with a closed 3-form $H$, $d H=0$. Then one may add a term to the corresponding action functional as 
\be 
S_{\text{WZPSM}}=S_{\text{PSM}}+\int_{\S_3}H(X)\,,
\ee   
where $\S_3$ is a membrane worldvolume whose boundary is $\S_2$ and $H(X):=X^{\ast}(H)$ is the pull-back of the 3-form to $\S_3$. It turns out that the twisted model is invariant under a modified set of gauge transformations provided that the following twisted Poisson condition holds:
\be \label{twistedPoisson}
\frac 12 [\mathcal{P},\mathcal{P}]_{\text{S}}=\langle H,\otimes^3\mathcal{P}\rangle\,
\ee 
where the contraction on the right-hand side is with respect to the first entry of each appearance of the bivector $\mathcal{P}$.  In local coordinates this means that the 3-vector on the left-hand side is obtained by raising each of the indices of the 3-form by one $\mathcal{P}$, in particular 
\be 
3\mathcal{P}^{li}\partial_{l}\mathcal{P}^{jk}=\mathcal{P}^{pi}\mathcal{P}^{qj}\mathcal{P}^{rl}H_{pqr}\,.
\ee 
Our aim in this section is to study the analog of this twisted Poisson structure in the case of the Jacobi sigma model. Although one might think that this is completely straightforward due to the Poissonization trick, we will see that a proper twisted Jacobi structure requires a cautious definition. 

Given that apart from $X^i$, we now also have the additional scalar field $\Phi$, there are two membrane terms one may add to the model. They are the pull-backs by the sigma model map $\hat X=(X,\Phi)$ of a 3-form $\hat H\in\Gamma(\bigwedge^3(T^*M\oplus \R))$ and a 2-form $\hat \Omega\in\Gamma(\bigwedge^2(T^*M\oplus \R))$ on the target space. Note that in principle one may also consider a second 2-form and a vector, and then add their corresponding terms on $\S_2$; however, we will argue that these terms belong to the same cohomology classes of $\hat H$ and $\hat \Omega$ respectively. Finally, we mention that unlike the Poisson sigma model, here the 3-form $\hat H$ is not assumed to be closed from the beginning; we examine its properties below.
According to the above, we consider the action functional 
\bea\label{wzjsm}
S_{\text{WZJSM}}=S_{\text{JSM}}+\int_{\Sigma_3}\frac{1}{6}\hat H_{ijk}(\hat X)\dd X^i\w\dd X^j\w\dd X^k+\int_{\Sigma_3}\frac{1}{2}\hat \Omega_{ij}(\hat X)\dd X^i\w\dd X^j\w\dd\Phi~.
\eea
For the time being, we do not specify further the dependence of the components $\hat H_{ijk}$ and $\hat \Omega_{ij}$ on the scalar field $\Phi$. Furthermore, we suggest the following extended infinitesimal gauge transformations for the 1-form fields, denoted as $\hat{\d}$, now including contributions from the components of the Wess-Zumino terms, 
\bea
\hat{\d} A_i&=&\d A_i-\sfrac 12e^{-\Phi} \left(\hat{H}_{ijk}(\Pi^{kl}\e_l+V^k\e_0)+\hat{\Omega}_{ij}V^l\e_l\right)(\dd X^j-e^{-\Phi}\Pi^{jm}A_m+e^{-\Phi}V^jA_0)+ \nn\\ 
&& +\sfrac 12e^{-\Phi} \hat{\Omega}_{ij}(\Pi^{jl}\e_l+V^j\e_0)(\dd \Phi-e^{-\Phi}V^mA_m)\,,
\\[4pt]
\hat{\d} A_0&=&\d A_0-\sfrac 12 e^{-\Phi}\hat{\Omega}_{jk}(\Pi^{kl}\e_l+V^k\e_0)(\dd X^j-e^{-\Phi}\Pi^{jm}A_m+e^{-\Phi}V^jA_0)\,, 
\eea
up to trivial gauge transformations. On the other hand, the transformations for the scalar fields $X^i$ and $\Phi$ remain unchanged. Note that the rightmost parentheses of each new term contain combinations starting with $\dd X^i$ and $\dd \Phi$; clearly these are not the derivatives $DX^i$ and $D\Phi$ and they do not transform covariantly themselves. 

Next we examine the gauge invariance of the action functional \eqref{wzjsm}. First, a necessary condition is that 
\be\label{dh}
\dd \hat H+\dd\hat\Omega\w\dd\Phi=0~.
\ee
Thus we observe that at face value the 3-form $\hat H$ is not closed.
Then the transformation of \eqref{wzjsm} becomes:
\bea
\nn \hat\delta S_{\text{WZJSM}}&=&\int_{\Sigma_2}e^{-2\Phi}\left(\frac{1}{2}\epsilon_0 A_i\w A_j\left([\Pi,V]_{\text{S}}+e^{-\Phi}\left(\hat H(\Pi,\Pi,V)-\hat\Omega(\Pi,V)\w V\right)\right)^{ij}+\right.\\
\nonumber &&+\frac{1}{2}\epsilon_k A_i\w A_j\left(V\w\Pi-\frac{1}{2}[\Pi,\Pi]_{\text{S}}-e^{-\Phi}\left(\hat H(\Pi,\Pi,\Pi)+\hat \Omega(\Pi,\Pi)\w V\right)\right)^{ijk}+\\
&&\left.+\epsilon_j A_0\w A_i\left([\Pi,V]_{\text{S}}+e^{-\Phi}\left(\hat H(\Pi,\Pi,V)-\hat\Omega(\Pi,V)\w V\right)\right)^{ij}\right).
\eea
Therefore, gauge invariance is achieved if and only if the following conditions hold:
\bea
&&[\Pi,\Pi]_{\text{S}}=2\left(V\w\Pi-e^{-\Phi}\hat H(\Pi,\Pi,\Pi)-e^{-\Phi}V\w\hat\Omega(\Pi,\Pi)\right)~,\label{twistedJacobi1}\\
&&[\Pi ,V]_{\text{S}}=e^{-\Phi}V\w\hat\Omega(\Pi,V)-e^{-\Phi}\hat H(\Pi,\Pi,V)~.\label{twistedJacobi2}
\eea
Moreover, the result does not change essentially if we include from the beginning additional terms on the worldsheet $\S_2$. Such terms could be the pull-back of a 2-form $B$ and a 1-form $C$ of the type 
\be 
\int_{\S_2}\frac 12 B_{ij}(\hat X)\dd X^i\w\dd X^j+C_i(\hat X)\dd X^i\w\dd \Phi\,.
\ee  
The only difference then would be that the above conditions hold using instead the modified by exact terms 3-form and 2-form 
\be 
\hat H\to \hat H+\dd B \quad \text{and} \quad \hat\Omega\to \hat\Omega+\dd C\,,
\ee 
which belong to the same cohomology class as $\hat H$ and $\hat\Omega$. 

As a simple check at this stage, we observe that setting $\Phi$ and $V$ to zero the twisted Poisson condition \eqref{twistedPoisson} is immediately recovered. 
Then one could be tempted to call \eqref{twistedJacobi1} and \eqref{twistedJacobi2} the twisted Jacobi structure. However, a puzzle arises by noticing that the left-hand side of these conditions is independent of $\Phi$, while the right-hand side contains $\Phi$ explicitly. To resolve this puzzle, we are led to specify the dependence of $\hat H$ and $\hat\Omega$  on $\Phi$ by refining their definitions as follows:
\bea 
\hat H(X,\Phi)=e^{\Phi}H(X) \quad \text{and}\quad \hat\Omega(X,\Phi)=e^{\Phi}\Omega(X)\,.
\eea
Taking into account Eq. \eqref{dh}, it turns out that $H$ is exact (and therefore closed), and in particular 
\be 
H=\dd \Omega\,.
\ee  
We conclude that the twisted Jacobi structure is appropriately defined as a triple $(\Pi,V,\Omega)$ of a bivector field, a vector field and a nonclosed 2-form such that the following two conditions hold,
\bea
\frac 12[\Pi,\Pi]_{\text{S}}&=&V\w\Pi- \dd\Omega(\Pi,\Pi,\Pi)-V\w\Omega(\Pi,\Pi)~,\label{twistedJacobi1b}\\ 
{[}\Pi ,V]_{\text{S}}&=&V\w\Omega(\Pi,V)-\dd\Omega(\Pi,\Pi,V)~.\label{twistedJacobi2b}
\eea
This is in agreement with the corresponding definition of Ref. \cite{twistedJacobi}, which we obtained here using a field-theoretic approach and gauge invariance. 
Moreover, these observations show that the action functional of the twisted Jacobi sigma model ends up being 
\bea\label{wzjsm2}
S_{\text{WZJSM}}=S_{\text{JSM}}+\int_{\Sigma_3}\frac{1}{2}e^{\Phi}\partial_i\Omega_{jk}(X)\dd X^i\w\dd X^j\w\dd X^k+\int_{\Sigma_3}\frac{1}{2}e^{\Phi} \Omega_{ij}(X)\dd X^i\w\dd X^j\w\dd\Phi~,\nn
\eea
and noting that $\dd e^{\Phi}=e^{\Phi}\dd\Phi$, the Wess-Zumino term is a total derivative and thus drops to the boundary. Then the resulting theory is simply 
\bea\label{wzjsm3}
S_{\text{WZJSM}}&=&\int_{\S_2}A_i\w\dd X^i+A_0\w\dd\Phi+\nn\\ &&+\,\sfrac 12e^{-\Phi} \Pi^{ij}(X)A_i\w A_j+e^{-\Phi}V^i(X)A_0\w A_i+e^{\Phi}\Omega_{ij}(X)\dd X^i\w\dd X^j\,.
\eea
We remark that although the Poisson structure is a special case of Jacobi, this is no longer strictly true for the corresponding twisted structures. Setting $V=0$ to \eqref{twistedJacobi1b} and \eqref{twistedJacobi2b}, one obtains a twisted Poisson structure, albeit for an exact 3-form only. In general though the 3-form does not have to be exact but only closed. This shows that if one wishes to recover the general twisted Poisson structure, the case should be distinguished already at the level of conditions \eqref{twistedJacobi1} and \eqref{twistedJacobi2}. Indeed, our previous argumentation for defining the twisted Jacobi structure assumed that $\Phi$ is nonvanishing. However, when $\Phi$ is zero, then \eqref{dh} yields $\hat H$ closed and with $V$ also vanishing  the aforementioned conditions result in the twisted Poisson structure as already noted before. 

\section{Conclusions and outlook}

Recent years have faced a lot of activity in the understanding of the interplay between generalizations of geometry and physics, especially in the framework of string theory. Two-dimensional nonlinear sigma models are a natural arena where this interplay takes place, due to the identification of their couplings with a symmetric and antisymmetric 2-tensor respectively, which may be combined in a so-called generalized metric. 
This is where interpolations of geometric structures, such as pre-symplectic and Poisson \cite{dirac} or complex and symplectic \cite{Hitchin:2004ut} become relevant. The symmetry structure of such generalized geometries goes beyond diffeomorphisms due to the presence of antisymmetric tensor fields and it is captured by generalizations of the Lie bracket of vector fields, notably by the Courant bracket. In the presence of a closed 3-form, representing the field strength of the 2-form and being identified with a Wess-Zumino term in the sigma model, these structures are twisted. However, the possibility of a 3-form that is not closed has not been studied in detail from this field-theoretical viewpoint. Nonclosed 3-forms may appear either through dimensional reduction of closed 3-forms or in the context of Heterotic string theory, where the Bianchi identity is modified due to $\a'$ corrections.

Motivated by the above, we have studied two main aspects of nonlinear sigma models with Wess-Zumino term. First, generalizing the approach of \cite{Salnikov:2013pwa,Chatzistavrakidis:2016jci,ChatzistavrakidisAHP}, we have studied the gauging of a class of sigma models where the Wess-Zumino term is not closed, a situation that can arise through dimensional reduction. We have shown that the constraints which guarantee consistency of the gauging procedure are directly associated with the structure of a  transitive Courant algebroid on the effective target space $TM\oplus {\cal G}\oplus T^{\ast}M$, where ${\cal G}$ is a bundle of quadratic Lie algebras. In case that ${\cal G}=M\times \R\oplus \R$, the constraints are identified with the conditions for isotropic and involutive subbundles of the fully twisted contact Courant algebroid of \cite{BouwknegtLectures}, namely contact Dirac structures.

Furthermore, motivated by the fact that the Poisson sigma model can result from a gauging procedure and moreover that it may be twisted by a 3-form in a way that defines a generalization of Poisson manifolds \cite{Klimcik:2001vg}, we have studied sigma models whose target space is endowed with a Jacobi structure. Jacobi structures are generalizations of Poisson in the presence of an additional vector field that controls the nonvanishing of the Schouten bracket of the would-be Poisson bivector with itself. In addition, they are related to Poisson via a Poissonization theorem stating that a Jacobi structure can be oxidized to a Poisson in one dimension higher. Utilizing this fact, we have constructed Jacobi sigma models and studied their gauge structure. Gauge invariance is controlled by the defining relations of this more general structure and the gauge algebra is an open one. Finally, we have included a Wess-Zumino term in the Jacobi sigma model and demonstrated that gauge invariance leads to a twisted Jacobi structure, which we identify.

The above results can be useful in the context of Heterotic string theory and its sigma model. Indeed, the Bianchi identity of the 3-form in Heterotic string theory is modified with respect to type II superstrings, in that it acquires $\a'$ corrections and thus is of the general form studied in this paper. It would be interesting to construct Heterotic membrane sigma models in analogy to the Courant sigma model \cite{Ikeda:2002wh}, with the expectation that the Heterotic Bianchi identity follows from the classical master equation in the BV-BRST formulation. This is expected also in view of analogous results in the context of QP manifolds \cite{Deser:2017fko}. One could then also expect to obtain a set of geometric and nongeometric fluxes for Heterotic string and double field theory following the strategy of \cite{Chatzistavrakidis:2018ztm}. A further important goal would then be to understand the geometry of $\a'$ corrections. Recently, the inclusion of such leading order corrections in the bracket and bilinear form of Courant algebroids was suggested \cite{Hohm:2013jaa,Hohm:2014eba,Bedoya:2014pma}, as well as their relation to transitive Courant algebroids \cite{Coimbra:2014qaa}. Moreover, \cite{Baron:2018lve} suggests a method to account for higher order corrections. It would be interesting to understand the geometric realization of this method in the context of higher Courant algebroids.         

Aside string theory, contact and Jacobi structures are relevant in the study of time-dependent and/or dissipative Hamiltonian systems \cite{chm}. Given that the origins of the Courant bracket are found in the study of constrained Hamiltonian mechanics, with the outcome that pre-symplectic and Poisson structures are essentially treated on an equal footing, it would be interesting to explore this spirit with reference to pre-contact and Jacobi structures too. This could find applications in classical time-dependent systems and their quantization.  

\paragraph{Acknowledgements.} We would like to thank Larisa Jonke, Zolt\'an K\"ok\'enyesi \& Thomas Strobl for useful comments and suggestions on the manuscript. A. Ch. also thanks Thomas Strobl for interesting discussions and the Erwin Schr\"odinger International Institute for Mathematics and Physics for hospitality and support during the programme ``Higher Structures and Field Theory''. This work is supported by the Croatian Science Foundation Project ``New Geometries for Gravity and Spacetime'' (IP-2018-01-7615), and also partially supported by the European Union through the European Regional Development Fund - The Competitiveness and Cohesion Operational Programme (KK.01.1.1.06).


\begin{thebibliography}{99}
	
	\bibitem{SchallerStrobl}
	P.~Schaller and T.~Strobl,
	``Poisson structure induced (topological) field theories,''
	Mod.\ Phys.\ Lett.\ A {\bf 9} (1994) 3129
	[hep-th/9405110].
	
	\bibitem{Ikeda}
	N.~Ikeda,
	``Two-dimensional gravity and nonlinear gauge theory,''
	Annals Phys.\  {\bf 235} (1994) 435
	[hep-th/9312059].
	
	\bibitem{CattaneoFelder1}
	A.~S.~Cattaneo and G.~Felder,
	``A Path integral approach to the Kontsevich quantization formula,''
	Commun.\ Math.\ Phys.\  {\bf 212} (2000) 591
	[math/9902090].
	
	\bibitem{CattaneoFelder2}
	A.~S.~Cattaneo and G.~Felder,
	``Poisson sigma models and deformation quantization,''
	Mod.\ Phys.\ Lett.\ A {\bf 16} (2001) 179
	[hep-th/0102208].
	
	\bibitem{Klosch:1995fi}
	T.~Klosch and T.~Strobl,
	``Classical and quantum gravity in (1+1)-Dimensions. Part 1: A Unifying approach,''
	Class.\ Quant.\ Grav.\  {\bf 13} (1996) 965
	Erratum: [Class.\ Quant.\ Grav.\  {\bf 14} (1997) 825]
	[gr-qc/9508020].
	
	\bibitem{Grumiller:2002nm}
	D.~Grumiller, W.~Kummer and D.~V.~Vassilevich,
	``Dilaton gravity in two-dimensions,''
	Phys.\ Rept.\  {\bf 369} (2002) 327
	[hep-th/0204253].
	
	
	\bibitem{Alekseev:1995py}
	A.~Y.~Alekseev, P.~Schaller and T.~Strobl,
	``The Topological G/G WZW model in the generalized momentum representation,''
	Phys.\ Rev.\ D {\bf 52} (1995) 7146
	[hep-th/9505012].
	
	\bibitem{Alexandrov:1995kv}
	M.~Alexandrov, A.~Schwarz, O.~Zaboronsky and M.~Kontsevich,
	``The Geometry of the master equation and topological quantum field theory,''
	Int.\ J.\ Mod.\ Phys.\ A {\bf 12} (1997) 1405
	[hep-th/9502010].
	
	\bibitem{Ikeda:2012pv}
	N.~Ikeda,
	``Lectures on AKSZ Sigma Models for Physicists,''
	arXiv:1204.3714 [hep-th].
	
	\bibitem{Kotov:2004wz}
	A.~Kotov, P.~Schaller and T.~Strobl,
	``Dirac sigma models,''
	Commun. Math. Phys. \textbf{260} (2005), 455-480
	[arXiv:hep-th/0411112 [hep-th]].
	
	 \bibitem{dirac}
	T.~Courant, 
	``Dirac manifolds,''
	Trans. Amer. Math. Soc., {\bf 319} (1990) 631-661. 
	
	\bibitem{Liu:1995lsa}
	Z.~Liu, A.~Weinstein and P.~Xu,
	``Manin Triples for Lie Bialgebroids,''
	J. Diff. Geom. \textbf{45} (1997) no.3, 547-574
	[arXiv:dg-ga/9508013 [math.DG]].
	
	\bibitem{gg1}
	M.~Gualtieri,
	``Generalized complex geometry,'' Oxford University DPhil Thesis, 
	math/0401221 [math-dg].
	
	\bibitem{hs1}
	C.~M.~Hull and B.~J.~Spence,
	``The Gauged Nonlinear $\sigma$ Model With {Wess-Zumino} Term,''
	Phys.\ Lett.\ B {\bf 232} (1989) 204.
	
	\bibitem{hs2}
	C.~M.~Hull and B.~J.~Spence,
	``The Geometry of the gauged sigma model with Wess-Zumino term,''
	Nucl.\ Phys.\ B {\bf 353} (1991) 379.
	
	\bibitem{Salnikov:2013pwa}
	V.~Salnikov and T.~Strobl,
	``Dirac sigma models from gauging,''
	JHEP \textbf{11} (2013), 110
	[arXiv:1311.7116 [math-ph]].
	
	\bibitem{Plauschinn:2013wta}
	E.~Plauschinn,
	``T-duality revisited,''
	JHEP \textbf{01} (2014), 131
	[arXiv:1310.4194 [hep-th]].
	
	\bibitem{Chatzistavrakidis:2016jci}
	A.~Chatzistavrakidis, A.~Deser, L.~Jonke and T.~Strobl,
	``Beyond the standard gauging: gauge symmetries of Dirac Sigma Models,''
	JHEP \textbf{08} (2016), 172
	[arXiv:1607.00342 [hep-th]].
	
	\bibitem{ChatzistavrakidisAHP}
	A.~Chatzistavrakidis, A.~Deser, L.~Jonke and T.~Strobl,
	``Strings in Singular Space-Times and their Universal Gauge Theory,''
	Annales Henri Poincare \textbf{18} (2017) no.8, 2641-2692
	[arXiv:1608.03250 [math-ph]].
	
	\bibitem{Severa:2019ddq}
	P.~\v{S}evera and T.~Strobl,
	``Transverse generalized metrics and 2d sigma models,''
	J. Geom. Phys. \textbf{146} (2019), 103509
	[arXiv:1901.08904 [math.DG]].
	
	\bibitem{Ikeda:2018rwe}
	N.~Ikeda and T.~Strobl,
	``On the relation of Lie algebroids to constrained systems and their BV/BFV formulation,''
	Annales Henri Poincare \textbf{20} (2019) no.2, 527-541
	[arXiv:1803.00080 [math-ph]].
	
	\bibitem{Ikeda:2019pef}
	N.~Ikeda,
	``Momentum sections in Hamiltonian mechanics and sigma models,''
	SIGMA \textbf{15} (2019), 076
	[arXiv:1905.02434 [math-ph]].
	
	\bibitem{BouwknegtLectures}
	P.~Bouwknegt,
	``Lectures on cohomology, T-duality, and generalized geometry,''
	Lect.\ Notes Phys.\  \textbf{807} (2010), 261-311
	
	\bibitem{Coimbra:2014qaa}
	A.~Coimbra, R.~Minasian, H.~Triendl and D.~Waldram,
	``Generalised geometry for string corrections,''
	JHEP \textbf{11} (2014), 160
	[arXiv:1407.7542 [hep-th]].
	
	\bibitem{Severa:2017oew}
	P.~\v{S}evera,
	``Letters to Alan Weinstein about Courant algebroids,''
	[arXiv:1707.00265 [math.DG]].
	
	\bibitem{Bursztyn:2005vwa}
	H.~Bursztyn, G.~R.~Cavalcanti and M.~Gualtieri,
	``Reduction of Courant algebroids and generalized complex structures,''
	Adv. Math. \textbf{211} (2007), 726-765
	[arXiv:math/0509640 [math.DG]].
	
	\bibitem{RCA}
	Z.~Chen, M.~Sti\'enon, P.~Xu, ``On regular Courant algebroids,'' J. Symplectic Geom. 11 (2013), no. 1, 1--24. 
	
	\bibitem{Severa:2015hta}
	P.~\v{S}evera,
	``Poisson-Lie T-Duality and Courant Algebroids,''
	Lett. Math. Phys. \textbf{105} (2015) no.12, 1689-1701
	[arXiv:1502.04517 [math.SG]].
	
	\bibitem{Wright}
	K.~Wright,
	``Generalised contact geometry as reduced generalised complex geometry,''
	J.\ Geom.\ Phys.\  \textbf{130} (2018), 331-348
	[arXiv:1708.09550 [math.DG]].
	
	\bibitem{ks1}
	A.~Kotov and T.~Strobl,
	``Gauging without initial symmetry,''
	J.\ Geom.\ Phys.\  {\bf 99} (2016) 184
	[arXiv:1403.8119 [hep-th]].
	
		\bibitem{Klimcik:2001vg}
		C.~Klimcik and T.~Strobl,
		``WZW - Poisson manifolds,''
		J.\ Geom.\ Phys.\  {\bf 43} (2002) 341
		[math/0104189 [math-sg]].
		
		\bibitem{SeveraWeinstein}
		P.~\v{S}evera and A.~Weinstein,
		``Poisson geometry with a 3 form background,''
		Prog.\ Theor.\ Phys.\ Suppl.\  {\bf 144} (2001) 145
		[math/0107133 [math-sg]].
		
	
	\bibitem{Ikeda:2019czt}
	N.~Ikeda and T.~Strobl,
	``BV and BFV for the H-twisted Poisson sigma model,''
	[arXiv:1912.13511 [hep-th]].

\bibitem{Jacobi} 
A. Lichnerowicz, 
``Les vari\'et\'es de Jacobi et leurs alg\'ebres de Lie associ\'ees,'' J. Differential Geometry 12 (1977), 253–300.

\bibitem{Crainic}
M.~Crainic, C.~Zhu,
``Integrability of Jacobi structures,''
Annales de l'Institut Fourier, Volume 57 (2007) no. 4, p. 1181-1216. 

\bibitem{Vaisman} 
I.~Vaisman, \textit{Lectures on the Geometry of Poisson Manifolds}, Birkh\"auser Verlag 1994


\bibitem{twistedJacobi}
J.~M.~Nunes da Costa and F.~Petalidou, 
``Twisted Jacobi Manifolds, Twisted Dirac–Jacobi Structures and Quasi-Jacobi Bialgebroids,'' 
Journal of Physics A: Mathematical and General 39.33 (2006): 10449–10475.

\bibitem{Witten:1983ar}
E.~Witten,
``Nonabelian Bosonization in Two-Dimensions,''
Commun. Math. Phys. \textbf{92} (1984), 455-472


\bibitem{Iglesias}
D.~Iglesias and J.C.~Marrero, ``Generalized Lie bialgebroids and Jacobi structures,'' Journal of Geometry and Physics 40 (2001), 176-200.  

\bibitem{GM}
J.~Grabowski, G.~Marmo, ``Jacobi structures revisited,'' Journal of Physics A: Mathematical and General, Volume 34, Number 49 (2001).

\bibitem{HT}
M.~Henneaux and C.~Teitelboim, ``Quantization of Gauge Systems'', Princeton University Press (1992).

\bibitem{Chatzistavrakidis:2017tpk}
A.~Chatzistavrakidis, A.~Deser, L.~Jonke and T.~Strobl,
``Gauging as constraining: the universal generalised geometry action in two dimensions,''
PoS \textbf{CORFU2016} (2017), 087
[arXiv:1705.05007 [hep-th]].

\bibitem{bourgeois} 
F.~Bourgeois,
``A survey of contact homology,''
in ``New Perspectives and Challenges in Symplectic Field Theory'', CRM Proceedings and Lecture Notes 49 (2009), 45-72

\bibitem{kirillov} 
A.~Kirillov, ``Local Lie Algebras,'' Russian Mathematical Surveys, Volume 31, N.~4 (1976). 

\bibitem{marle}
C.-M.~Marle, 
``On Jacobi Manifolds and Jacobi Bundles,'' In: Dazord P., Weinstein A. (eds) Symplectic Geometry, Groupoids, and Integrable Systems. Mathematical Sciences Research Institute Publications, vol 20. Springer, New York (1991).

\bibitem{Witten:1988xj}
E.~Witten,
``Topological Sigma Models,''
Commun. Math. Phys. \textbf{118} (1988), 411.

\bibitem{Hitchin:2004ut}
N.~Hitchin,
``Generalized Calabi-Yau manifolds,''
Quart. J. Math. \textbf{54} (2003), 281-308
[arXiv:math/0209099 [math.DG]].

\bibitem{Ikeda:2002wh}
N.~Ikeda,
``Chern-Simons gauge theory coupled with BF theory,''
Int. J. Mod. Phys. A \textbf{18} (2003), 2689-2702
[arXiv:hep-th/0203043 [hep-th]].


\bibitem{Deser:2017fko}
A.~Deser, M.~A.~Heller and C.~S\"amann,
``Extended Riemannian Geometry II: Local Heterotic Double Field Theory,''
JHEP \textbf{04} (2018), 106
[arXiv:1711.03308 [hep-th]].

\bibitem{Chatzistavrakidis:2018ztm}
A.~Chatzistavrakidis, L.~Jonke, F.~S.~Khoo and R.~J.~Szabo,
``Double Field Theory and Membrane Sigma-Models,''
JHEP \textbf{07} (2018), 015
[arXiv:1802.07003 [hep-th]].

\bibitem{Hohm:2013jaa}
O.~Hohm, W.~Siegel and B.~Zwiebach,
``Doubled $\alpha'$-geometry,''
JHEP \textbf{02} (2014), 065
[arXiv:1306.2970 [hep-th]].

\bibitem{Hohm:2014eba}
O.~Hohm and B.~Zwiebach,
``Green-Schwarz mechanism and $\alpha'$-deformed Courant brackets,''
JHEP \textbf{01} (2015), 012
[arXiv:1407.0708 [hep-th]].

\bibitem{Bedoya:2014pma}
O.~A.~Bedoya, D.~Marques and C.~Nunez,
``Heterotic $\alpha$'-corrections in Double Field Theory,''
JHEP \textbf{12} (2014), 074
[arXiv:1407.0365 [hep-th]].

\bibitem{Baron:2018lve}
W.~H.~Baron, E.~Lescano and D.~Marques,
``The generalized Bergshoeff-de Roo identification,''
JHEP \textbf{11} (2018), 160
[arXiv:1810.01427 [hep-th]].

\bibitem{chm}
Alessandro Bravetti, Hans Cruz, Diego Tapias,
``Contact Hamiltonian mechanics,''
Annals of Physics \textbf{376} (2017), 17-39.

\bibitem{jsm}
F. Bascone, F. Pezzella, P. Vitale, ``Jacobi Sigma Models'', arXiv: 2007.12543 [hep-th].


\end{thebibliography}
\end{document}